\documentclass[journal,twocolumn]{IEEEtran}

\usepackage{xcolor}
\usepackage{multirow}
\usepackage{graphicx}
\usepackage{subfigure}
\usepackage{algorithm}
\usepackage{algorithmicx}
\usepackage{algpseudocode}
\usepackage{ifthen}
\usepackage[figuresright]{rotating}
\usepackage{amsmath}
\usepackage{amssymb}
\usepackage{amsthm}
\usepackage{booktabs}
\usepackage{cite}
\usepackage{bm}
\usepackage{multicol} 
\usepackage{enumerate}
\usepackage{url}
\usepackage[hidelinks]{hyperref}
\usepackage[flushleft]{threeparttable}
\usepackage{float}
\usepackage{stfloats}
\usepackage{CJKutf8}
\usepackage{marvosym}
\usepackage{ifsym}
\usepackage{upgreek}

\begin{document}

\title{HMAMP: Hypervolume-Driven Multi-Objective Antimicrobial Peptides Design}

\author{Li~Wang, Yiping~Liu~\Letter, Xiangzheng Fu~\Letter, Xiucai~Ye, Junfeng Shi, Gary G. Yen, Fellow, IEEE, Xiangxiang Zeng
\thanks{Corresponding Authors: Yiping Liu (yiping0liu@gmail.com), Xiangzheng Fu(excelsior511@126.com); Other Authors:Li~Wang(wl12345678@hnu.edu.cn), Xiucai~Ye(yexiucai@cs.tsukuba.ac.jp) Junfeng~Shi(Jeff-shi@hnu.edu.cn), Gary G. Yen(gyen@okstate.edu), Xiangxiang Zeng 
(xzeng@foxmail.com)} 
\thanks{This work was supported by the National Natural Science Foundation of China (Grant No. 662002111,  62372158, 62106073, 62122025, U22A2037, 62250028), the Natural Science Foundation of Hunan Province (Grant No. 2022JJ40090))} 

\thanks{
Li~Wang, Yiping~Liu, X. Zeng, and Junfeng Shi are with the College of Computer Scienceand Electronic Engineering, Hunan University, Changsha, Hunan,
(wl12345678@hnu.edu.cn, yiping0liu@gmail.com, xzeng@foxmail.com, Jeff-shi@hnu.edu.cn)

Gary G. Yen is with the School of Electrical and Computer Engineering, Oklahoma State University, Stillwater. OK 74078. USA. (gyen@okstate.edu)

Xiucai~Ye is with the School of System Information and Engineering, University of Tsukuba,Ibaraki, Japan(yexiucai@cs.tsukuba.ac.jp)

Xiangzheng Fu is with Neher's Biophysics Laboratory for Innovative Drug Discovery, State Key Laboratory of Quality Research in Chinese Medicine, Macau Institute for Applied Research in Medicine and Health, Macau University of Science and Technology ,Macau, China, 999078 (e-mail:excelsior511@126.com).
}
}

\markboth{  }%
{Shell \MakeLowercase{\textit{et al.}}: Bare Demo of IEEEtran.cls for IEEE Journals}

\maketitle

\begin{abstract}
Antimicrobial peptides (AMPs) have exhibited unprecedented potential as biomaterials in combating multidrug-resistant bacteria. Despite the increasing adoption of artificial intelligence for novel AMP design, challenges pertaining to conflicting attributes such as activity, hemolysis, and toxicity have significantly impeded the progress of researchers. This paper introduces a paradigm shift by considering multiple attributes in AMP design.

Presented herein is a novel approach termed Hypervolume-driven Multi-objective Antimicrobial Peptide Design (HMAMP), which prioritizes the simultaneous optimization of multiple attributes of AMPs. By synergizing reinforcement learning and a gradient descent algorithm rooted in the hypervolume maximization concept, HMAMP effectively expands exploration space and mitigates the issue of pattern collapse. This method generates a wide array of prospective AMP candidates that strike a balance among diverse attributes. Furthermore, we pinpoint knee points along the Pareto front of these candidate AMPs. Empirical results across five benchmark models substantiate that HMAMP-designed AMPs exhibit competitive performance and heightened diversity. A detailed analysis of the helical structures and molecular dynamics simulations for ten potential candidate AMPs validates the superiority of HMAMP in the realm of multi-objective AMP design. The ability of HMAMP to systematically craft AMPs considering multiple attributes marks a pioneering milestone, establishing a universal computational framework for the multi-objective design of AMPs.

\end{abstract}


\IEEEpeerreviewmaketitle

\section{Introduction}
\IEEEPARstart{D}{eep} generative models show promising results when applied to material and drug discovery\cite{77,78,79,80,81}. They are also one of the viable ways to boost the speed of antimicrobial peptides(AMP) discovery, and plenty of works have been done to employ deep generative models in searching for AMP with desired properties and achieved great success\cite{37,38,39,40,41}. Given the massive peptide space, however,it is very likely that numerous AMPs are yet to be found.

\begin{figure}
\centering 
\includegraphics[width=1\columnwidth]{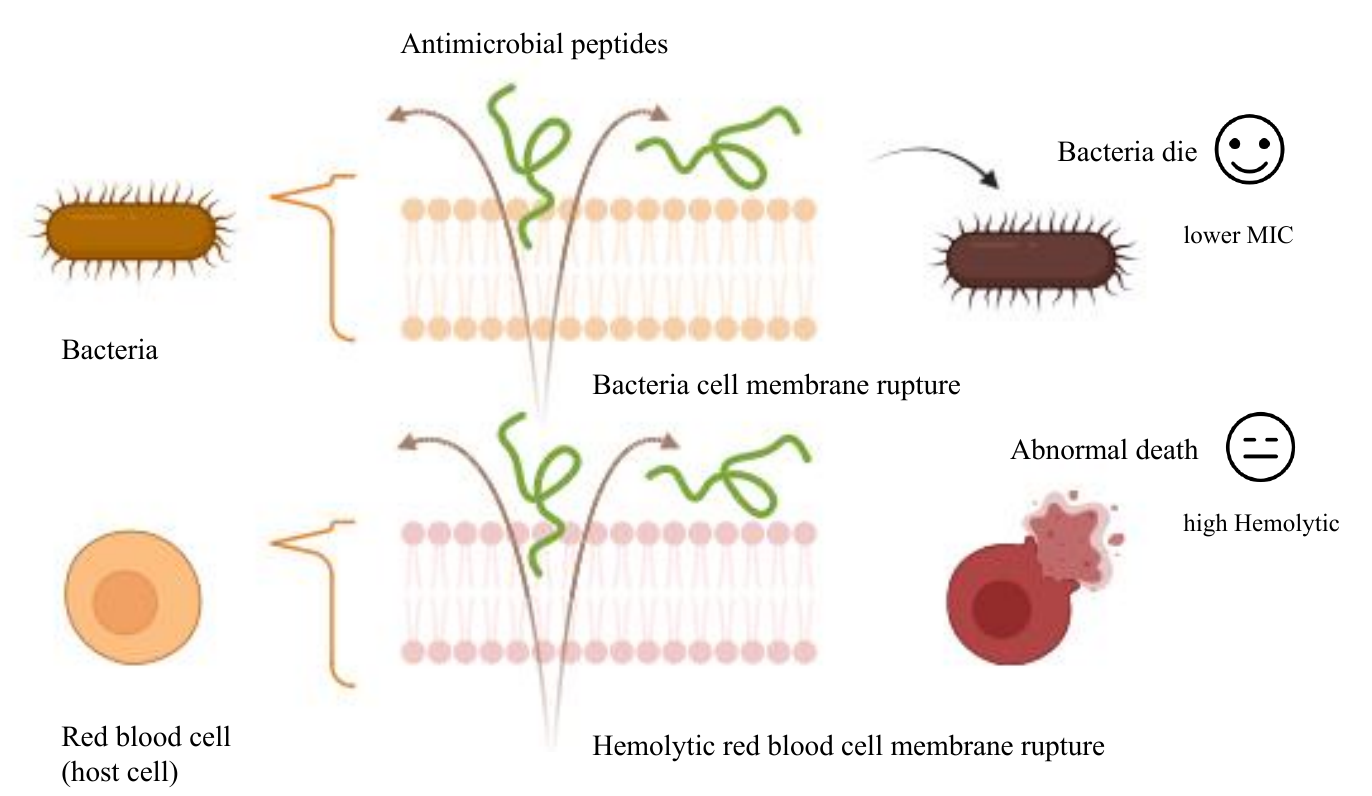}
\caption{Antimicrobial peptides(AMPs) cross cell membranes and act on bacteria and red blood cells, and low MIC and hemolysis conflict.}
\label{bater}
\end{figure}

AMPs are short (10-100 amino acid) peptides that possess an overall positive charge and a large percentage of hydrophobic amino acids\cite{83,84}, characterizing by various properties, each informative of its clinical potential including activity, toxicity, and hemolysis. Activity is measured in antimicrobial assays against various bacterial strains as minimum inhibitory concentration (MIC)\cite{82}. The most prominent AMPs have low values of MIC, implying that they remain active even in low concentrations, but their prevalence is limited; Toxicity is examined across diverse cell types (fibroblast, colon, lung, and cancer lines)\cite{85}; Hemolysis indicates the peptide concentration causing 50\% hemolysis in human or other mammals' (rabbits, sheep) erythrocytes\cite{86}. However, when the MIC value surpasses a certain threshold, the potential for detriment to the integrity of red blood cell membranes emerges, leading to the unwished hemolysis (as shown in Figure \ref{bater}). In addition, highly active antimicrobial peptides may not only be harmful to pathogens, but also have certain toxicity to host cells. Therefore, simultaneously optimizing multiple attributes of AMPs is a conflicting and challenging task.

Nonetheless, conventional methodologies for isolating superior AMPs from diverse organisms are both time-intensive and costly. As a result, the demand for efficient and precise computer simulation techniques has grown, aimed at generating novel AMPs to expedite the process of discovering these potent antimicrobial agents.

The pursuit of discovering peptide sequences with specific desired attributes primarily fall into two distinct categories: the exploration of novel AMPs from naturally occurring sequences, and the deliberate design of artificial AMPs through the modification of known ones or by originating them from the ground up. The strategy of discovering AMPs involves prognosticating potential peptides with desired attributes by virtually screening extensive libraries of established peptides. However, this approach is constrained by its reliance on existing data and lacks exploration of novel peptide space, thus conferring a degree of dependency.

On the other hand, AMP design methodologies have witnessed robust advancement and validation for their efficacy. Examples include the creation of AMPs from scratch utilizing genetic algorithms\cite{10}, long short-term memory networks (LSTMs)\cite{37}, generative adversarial networks(GANs) \cite{74}, and autoencoders - AEs \cite{40}), Monte Carlo tree search\cite{13}, reinforcement learning (RL)\cite{14}, transfer learning \cite{15}, and Bayesian optimization\cite{16}.

Despite the promising outcomes derived from these methodologies, a critical challenge remains: the simultaneous optimization of multiple attributes of AMPs. The existing approaches often encounter conflicting and limited trade-offs between various attributes. The resulting candidate AMPs tend to enhance one specific attribute, while potentially disregarding other vital attributes. Consequently, AMP discovery or designs, as addressed by numerous studies, lacks specificity and utility for AI-driven AMP discovery.

\begin{figure}
\centering 
\includegraphics[width=1\columnwidth]{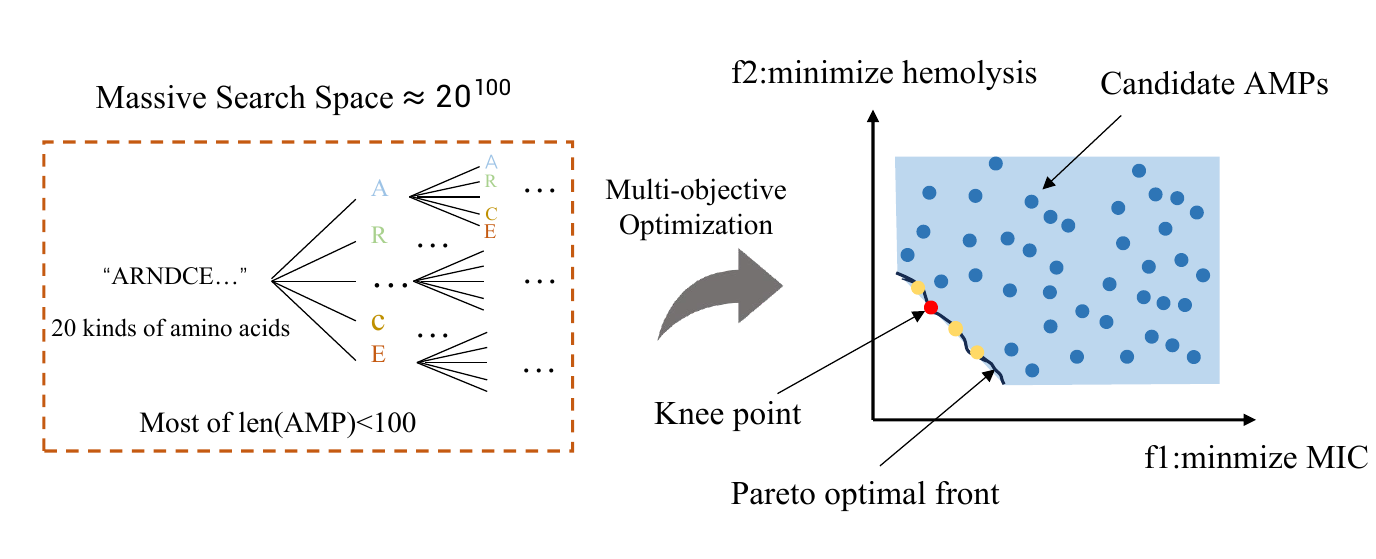}
\caption{The challenges we are attempting to address can be summarized as follows: Firstly, AMPs are composed of 20 different amino acids with a sequence length of less than 100, resulting in an enormous sequence space. Secondly, there are apparent conflicts in the properties of AMPs, and our objective is to simultaneously minimize multiple attributes (such as MIC, and hemolysis) to generate a balanced Pareto-optimal set considering various properties of AMPs. Finally, it is crucial to identify solutions on the optimal Pareto frontier that are of particular interest to decision-makers, such as the most promising knee points.}
\label{space}
\end{figure}

The task of AMP design can be conceptualized as a multi-objective optimization problem, involving the simultaneous enhancement of multiple conflicting attributes of AMPs. This study introduces a multi-objective antimicrobial peptides design using a Hypervolume-driven Multi-objective generative model (HMAMP). This approach first employs a multi-objective reinforcement learning generative network, which serves as the foundation for generating a collection of candidate AMPs. Then, the Pareto front of these candidate AMPs are formed by multiple deep-learning-based predictors. Subsequently, the Kneedle algorithm identifies knee points on this Pareto optimal front, representing the most captivating and promising AMP candidates. Following the above-mentioned technique, we will address the optimization challenges of antimicrobial peptides. The preliminary design is depicted in Figure \ref{space}.

To illustrate, considering a two-objective scenario, we formulate dual-attribute discriminators dedicated to comprehending the MIC and hemolytic attributes intrinsic to AMPs. By means of adversarial training within a Generative Adversarial Network (GAN) framework, the generation process is controlled to yield peptides that align with the designated attributes. Notably, the concepts of reinforcement learning and hypervolume maximization are synergistically employed to amplify exploration across a broader solution space and to foster training stability.

To ascertain the biological activity of the generated AMPs, we refine the attribute prediction of two predictors tailored for AMPs, leveraging the Pro-BERT-BFD framework. This refinement is employed to generate Pareto optimal AMPs featuring multifaceted objectives. Subsequently, the top ten AMP candidates are meticulously assessed, and the knee point among them is identified as the ultimate optimal AMPs. By subjecting these candidates to helical analysis and molecular dynamics simulations, we substantiate the exceptional activity and minimal side effects of these AMPs, affirming their pharmacokinetic viability.

In summary, this paper presents several significant contributions:

(1) Novel multi-objective AMPs generation model: We propose a novel approach for generating AMPs with desired multi-objective attributes. By merging generative adversarial and reinforcement learning techniques, our model addresses critical limitations seen in conventional approaches, such as restricted exploration space, instability, and susceptibility to pattern collapse.

(2) Hypervolume maximization: Incorporating the concept of Hypervolume maximization via gradient descent during training introduces a groundbreaking paradigm in AMP generation. This approach significantly improves solution set convergence and distribution, leading to enhanced quality and diversity of the generated AMPs.

(3) Enhanced screening and decision making on candidate AMPs: We meticulously train multiple predictors to forecast various attributes of the generated AMPs. This enables the creation of a Pareto front, leading to the selection of the most promising knee solutions. Leveraging the Kneedle algorithm for this selection process eliminates the need for laborious manual screening and decision making.

(4) Empirical Validation of HMAMP: Our experimental results underscore the prowess of HMAMP in crafting AMPs with desirable properties, including alpha helix formation, positive charge, favorable hydrophobicity, and effective amphiphilicity. These attributes hold paramount significance in enhancing the drug attributes of peptides for efficacious therapeutic applications.

Collectively, this paper's contributions pave the way for an advanced paradigm in AMP design, offering a comprehensive approach that transcends existing limitations and opens new avenues for multi-objective attribute optimization in drug development.

The remainder of the paper is structured as follows: In Section 2, we provide an in-depth review of the related work in the field of antimicrobial peptide design, highlighting the advancements, challenges, and gaps in existing methodologies. Section 3 outlines the proposed HMAMP framework, detailing its key components and the rationale behind its design. Subsequently, in Section 4, we expound upon the experimental setup and methodology, elucidating how the HMAMP approach was evaluated and its efficacy demonstrated. The obtained results and their implications are also discussed comprehensively. Finally, in Section 5, we offer conclusive remarks, summarizing the contributions of the work, highlighting its significance, and suggesting potential directions for future research endeavors in this domain. In supplementary material, we present definitions related to multi-objective optimization(see supplementary S1-S3) and GANs (see supplementary S4) from previous literature, which will prove useful in the subsequent sections.

\section{RELATED WORK}

In recent years, the search for known or predicted peptide sequences with the desired properties has become very popular, and the corresponding 
approaches are constantly being developed. Here, advanced computational strategies are presented and three groups of research approaches are distinguished: First, the discovery of new AMPs by prediction from naturally occurring sequences; 
Second, the batch design of novel AMPs from 
scratch; Third, the optimization design of protential AMPs focuses on multi-attribute constraint.

\subsection{AMPs Discovery by Prediction} 

Traditional approaches to AMPs design have heavily relied on structure-activity studies, often manifested as quantitative structure-activity relationship(QSAR) models \cite{17,18}. Such methods involve training machine learning models by identifying pertinent attributes, and subsequently utilizing these models to predict potential peptide candidates of interest. In more recent times, Ahmad et al. \cite{60} introduced a deep learning approach named Deep-AntiFP, comprising three dense layers, to predict antifungal peptides. This method uniquely investigated three distinct coding features of antifungal peptides and demonstrated superior performance when compared to existing models.

Furthermore, Ma et al. \cite{61} devised a DL technique that amalgamates predictions from five disparate DL models. This amalgamation was employed to discern AMPs from the human gut microbiome, employing peptide sequences as textual input and applying natural language processing methodologies to construct a unified pipeline. In another vein, Witten et al. \cite{62} amalgamated AMP classification with the regression of MIC, yielding an innovative approach for combating Escherichia coli, Pseudomonas aeruginosa, and Staphylococcus aureus.

However, it is essential to note that these QSAR-based methodologies are primarily geared towards scoring existing peptides and are not inherently equipped for the direct generation of new peptides.

\subsection{AMPs Design from scratch}

In the preliminary stages, the de novo design of new AMPs frequently incorporates the utilization of deep learning methods akin to those employed in natural language processing \cite{63,75}. For instance, Müller et al. \cite{37} harnessed a fusion of Long Short-Term Memory(LSTM) and Recurrent Neural Networks (RNN) to devise a linear cationic peptide endowed with an amphiphilic helix, thereby ensuring potent activity of the generated AMPs. An additional DL architecture, known as variational autoencoders (VAEs), has gained traction in the creation of novel chemical spaces \cite{64}. Szymczak et al. \cite{65} adopted AMP activity as an autonomous condition, proposing a low-dimensional continuous peptide feature space methodology rooted in conditional VAEs. This approach facilitated the generation of AMPs manifesting antimicrobial properties and robust activity.

In recent years, the ascendancy of GAN networks has been conspicuous in various generative tasks \cite{66}. Notably, Tucs et al. \cite{39} introduced PepGAN, a method utilizing GANs, with the objective of striking a balance between encompassing active peptides while concurrently sidestepping inactive ones. This dynamic culminated in the synthesis of the six most promising peptides. Although numerous methodologies are capable of generating AMPs with potent antimicrobial activity, these peptides often confront challenges in their translation to viable therapeutic drugs. Importantly, considerations regarding the stability, hemolysis, and toxicity of AMPs cannot be disregarded.

\subsection{AMPs Design by Optimization}

Among the early methods of optimized peptide design, a pivotal approach revolved around native peptide sequences. This technique engendered the creation of fresh, akin peptides by strategically removing or altering crucial residues within the peptide sequence. Taking cues from natural evolution, Yoshida et al. \cite{67} embarked on an exploration of the sequence space through a confluence of evolutionary algorithms and machine learning methodologies. Their endeavor entailed the design of novel AMPs commencing from template peptides. Over three iterative rounds of experimentation, they succeeded in achieving an astounding 160-fold enhancement in antimicrobial activity. It is imperative to note, however, that this work predominantly optimized analogs of the original sequence. Moreover, the employed mutations were informed solely by amino acid statistical data, leading to the exclusion of vast swathes of the sequence space.

Most recently, Hoffman et al. \cite{76} introduced a VAE-based model with
gradient descent zeroth-order optimization to convert a toxic peptide into a nontoxic one, while maintaining antimicrobial properties via similarity, which actually not targeted simultaneous enhancement of AMP properties. Liu et al. \cite{41} introduced an evolutionary multi-objective algorithm that harnessed de novo AMP generation, with antimicrobial activity and diversity as the focal optimization objectives. Although this methodology is adept at exploring a more extensive solution space, due to conflicting objectives,its consideration of the multi-pharmacokinetic properties of AMPs remains markedly constrained.

Moreover, relatively few methods account for the fact that each peptide should be optimized in not one, but several properties, e.g. be simultaneously active and nonhemolytic. In the present study, we present a pioneering approach named Hypervolume-driven Multi-objective Antimicrobial Peptide adversarial generative model (HMAMP). This innovative model undertakes the intricate task of learning the latent space housing diverse conflicting attribute peptides. Moreover, HMAMP effectively filters out AMPs featuring favorable drug properties via the Pareto front and knee points, thereby facilitating their application as efficacious treatments.

\section{Methods}

\begin{figure*}
\centering 
\includegraphics[width=\textwidth]{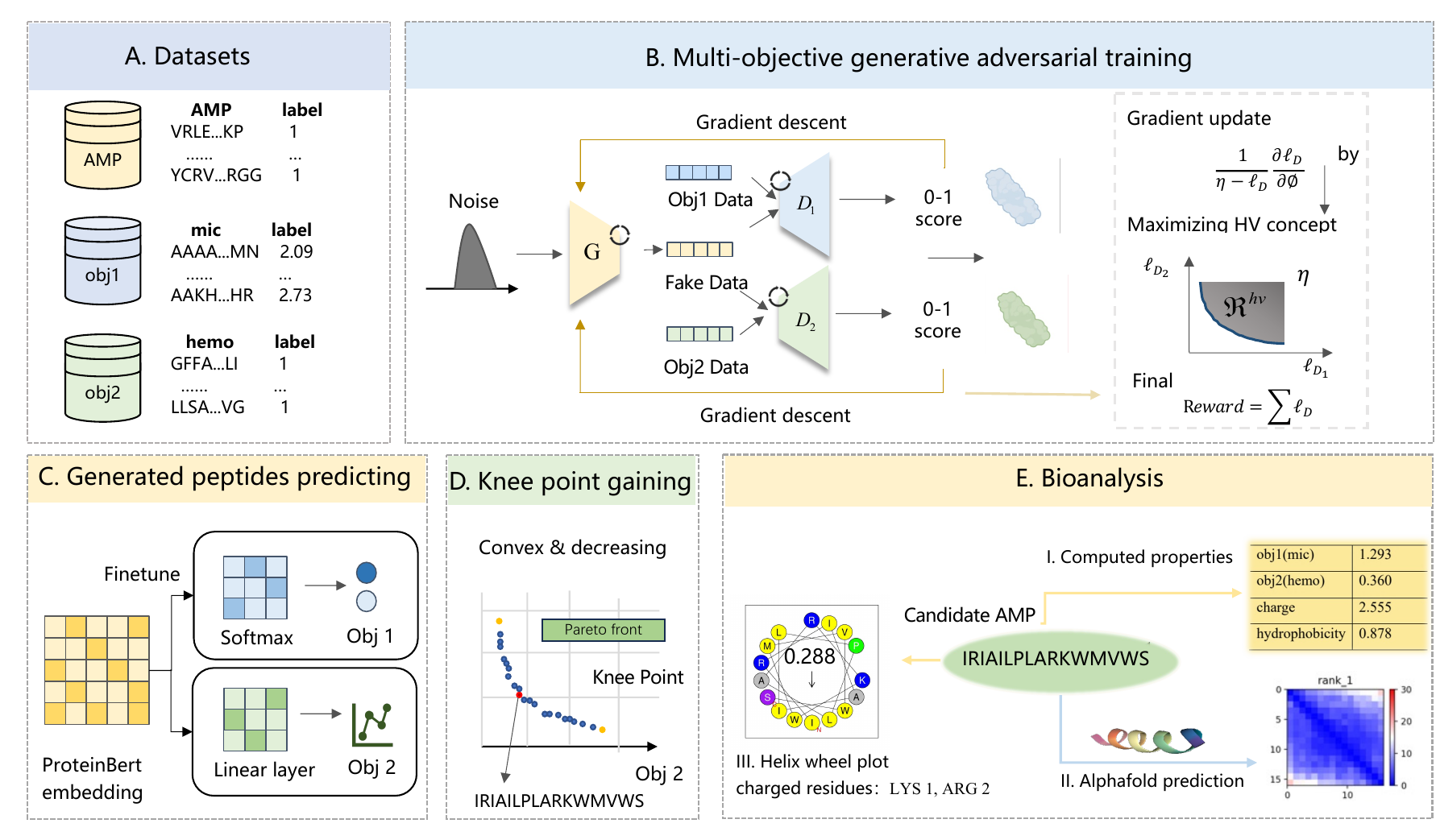}
\caption{Overview of the proposed HMAMP (shown here as a two-discriminator setup). Block(A) depicts the dataset of AMPs used for pretraining the generative model, with two additional target datasets contributing to the training of discriminators and predictors. Block(B) illustrates the application of the hypervolume maximization concept for training a multiple-discriminator generative model. Block(C) demonstrates the capabilities of the fine-tuned classifier and discriminator, achieved through Pro-BERT-BFD-based fine-tuning, for predicting the generated candidate AMPs. Block(D) displays the Pareto front of the generated AMPs along with the most promising knee point solutions, representing optimal AMP candidates. Block(E) portrays the AMPs module, subject to screening through helical structure analysis and molecular dynamics simulation, to validate the physicochemical attributes and secondary structure of the candidate AMPs.}
\label{HMAMP}
\end{figure*}
In this section, we elaborate the details of HMAMP. First, we employ multi-discriminator GAN for the multi-objective AMP design task to learn the multi-attribute latent space (Section \ref{section A}). Then, we leverage multi-objective optimization techniques, hypervolume maximization concept, to enhance the stability of generator training (Section \ref{section B}). Besides, we design two fine-tuning of multiple attribute predictors on the designated dataset correlated with candidate AMPs (Section \ref{section C}). Ultimately, driven by the outcomes of these predictors, we obtain the Pareto front inherent to the collection of candidate AMPs. From within this pareto front, we acquire the most captivating knee points through the utilization of the kneedle algorithm (Section \ref{section D}). The overview of our model is shown in Figure \ref{HMAMP}.

\subsection{AMP Generation Model via Multi-Objective GAN} \label{section A}

The AMP generator can be conceptualized as a stochastic policy within a reinforcement learning framework, which subsequently undergoes network updates using policy gradients \cite{26}. Reinforcement learning tackles the challenge of interacting with an environment, receiving rewards for actions, and formulating optimal strategies. This problem is often formulated as a Markov decision process, where environmental states $s\in S$ and agent actions $a\in A$ define an unknown environment with transition probabilities $p(s^{\prime}|s,a)$.

The agent's selection of action $a$ is contingent upon the conditional probability distribution $\pi_{\theta}(a|s)$, which is parameterized by $\theta\in\Theta$. The execution of this policy over the Markov decision process yields a trajectory $x=\{s_{t},a_{t}\}_{t=0}^{T}$, representing an AMP. The policy is then trained to maximize the expected reward over all possible trajectories $x$:

\begin{equation}
\overline{R} = E_{x \sim p_z}\left[R(x)\right] = \sum_x R(x) p_z(x)
\end{equation}

The reward function $R(x)$ can be defined as the negative summation of the individual discriminator losses:

\begin{equation}
\begin{aligned}
R(x) = -\sum\limits_{j=1}^k \mathcal{\ell}_{D_j} = \sum\limits_{j=1}^k \mathbb{E}_{\mathbf{x} \sim P_{\text{data}}}[\log(D_j(s,a))] + \\
\mathbb{E}_{\mathbf{z} \sim P_{\mathbf{z}}}[\log(1-D_j(s,a))]
\end{aligned}
\end{equation}

To address the multi-objective conflict generation challenge of AMPs, our AMP generator employs multiple discriminators. These discriminators facilitate the learning of an approximate distribution capable of effectively encompassing multi-modal data. In essence, for a given sample $x$ within the data space, the reward assigned to $x$ is high if it originates from $p_{\text{data}}$ within the data distribution, and low if extracted from sample distribution $p_z$. Unlike traditional GANs \cite{25}, our discriminators yield a reward value instead of a probability within the [0, 1] range. Additionally, the multi-discriminators share parameters. Formally, denoting the discriminators as $D_1, D_2, ..., D_k$, and the generator as $G$, we engage in a multiplayer minimax optimization game as follows:

\begin{equation}
\begin{aligned}
\min_G \max_{D_1, ..., D_k} \mathcal{J}(G, D_1, ..., D_k) &= \mathbb{E}_{\mathbf{x} \sim P_{\text{data}}}[\log D_1(\mathbf{x})] \\
&+ \mathbb{E}_{\mathbf{z} \sim P_{\mathbf{z}}}[-D_1(G(\mathbf{z}))] \\
&+ ... + \mathbb{E}_{\mathbf{x} \sim P_{\text{data}}}[-D_k(\mathbf{x})] \\
&+ \mathbb{E}_{\mathbf{z} \sim P_{\mathbf{z}}}[\log D_k(G(\mathbf{z}))].
\end{aligned}
\end{equation}

In this equation, $k$ represents the number of discriminators, with the current paper utilizing a setting of $k = 2$, focusing on attributes such as hemolysis and MIC. The discriminator endeavors to differentiate genuine trajectories from synthetic ones. Conversely, the generator seeks to perform optimally under the reward structure defined by the discriminator. This objective drives the generator to generate synthetic trajectories that are challenging to differentiate from the authentic generator ones, thereby "tricking" the discriminator.

\subsection{Hypervolume Maximization for Training Generator}\label{section B}

\begin{figure}
\centering 
\includegraphics[width=1\columnwidth]{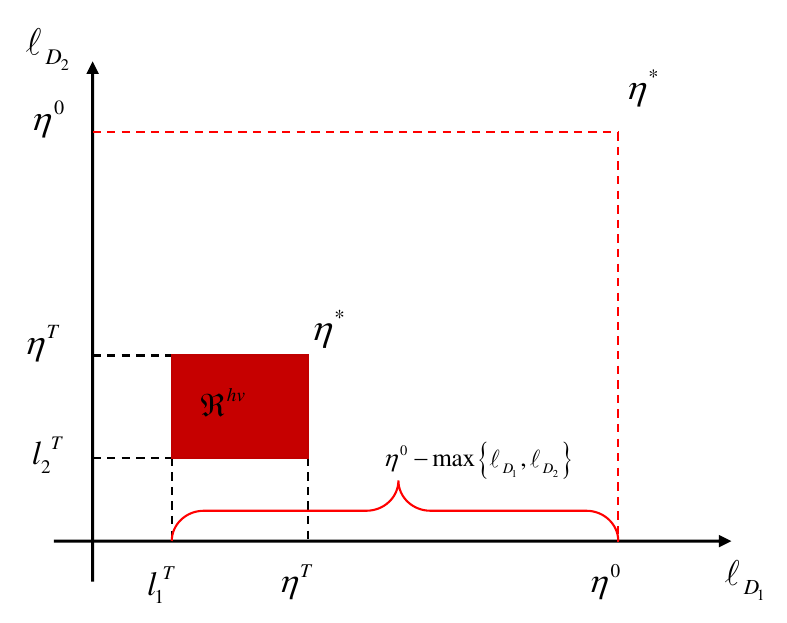}
\caption{2D example of the objective space where the generator loss is being optimized.}
\label{gradient}
\end{figure}

As mentioned earlier, we treat each discriminator's loss signal as an independent objective function and update the network using reinforcement learning's strategy gradient. To improve the stability of generator training, we leverage multi-objective optimization techniques to minimize losses, as illustrated in Figure \ref{gradient}. A well-established approach is Multi-Gradient Descent (MGD), but for large neural networks, MGD's computational cost becomes prohibitive. We recommend adopting a more efficient method that optimize by maximizing the Hypervolume (HV) concept. In contrast to MGD, the HV-maximization-based method places greater emphasis on feedback from underperforming discriminators.

We formulate the generator loss as the negative logarithm of the hypervolume, defined in Eq. \ref{eq9}:

\begin{equation}\label{eq9}
\mathcal{L}_G = -\mathfrak{R}^{h \nu} = -\sum_{j=1}^k \log(\eta^* - \mathcal{\ell}_{D_j}),
\end{equation}
where the reference point coordinate $\eta$ is an upper bound for all $\mathcal{\ell}_{D_j}$, and $\delta$ is a user-defined slack factor ensuring that $\eta$ remains adaptable, usually slightly above 1. As the reference point $\eta$ is held constant, $\mathfrak{R}^{h \nu}$ is maximized, consequently minimizing $\mathcal{L}_G$, when each $\mathcal{\ell}_{D_j}$ is minimized. The loss gradient for each discriminator is computed as follows:

\begin{equation}\label{eq11}
\frac{\partial\mathcal{L}_G}{\partial\phi} = \sum_{j=1}^k \frac{1}{\eta - \ell_{D_j}} \frac{\partial \ell_{D_j}}{\partial\phi}.
\end{equation}

This formulation places a stronger emphasis on higher losses in the final gradient, which is another advantage of employing the HV maximization optimization method.

\subsection{Multi-Attribute Predictors Learning}\label{section C}

Transfer learning has emerged as a pivotal technique in the realm of deep learning model development. It entails training a model on a source dataset and then transferring and fine-tuning it on a target domain \cite{27,28}. The Prot-BERT-BFD model \cite{68}, a transformer-based architecture boasting over 400 million parameters, is trained on a vast 2 billion protein fragments. The extensive protein sequence knowledge it acquires renders it highly applicable for peptide classification. In our study, we harness transfer learning from the Prot-BERT-BFD model to the target attribute model, aiming to impart the wealth of knowledge amassed by the former to the latter.

We establish both a Minimum Inhibitory Concentration (MIC) prediction model and a hemolytic classification model. Subsequently, these two attribute predictors undergo training and evaluation on a partitioned dataset with a distribution ratio of 6:2:2. The attained accuracy on an independent validation set surpasses 95\%. By leveraging these trained predictors, we predict the attribute values for candidate AMPs. This process facilitates the construction of an attribute target space, leading to the acquisition of the Pareto front for generated AMPs within this space. AMPs on the Pareto front are selected based on the concept of Pareto dominance in multi-objective optimization. This procedural step facilitates the preliminary screening of candidate AMPs.

\subsection{Knee Points Screening}\label{section D}

The Pareto front encompasses all non-dominated solutions to a multi-objective optimization problem, indicating that enhancing any single objective would result in the deterioration of other objectives. However, the Pareto front still contains numerous candidate AMPs. To enhance the selection of meaningful AMP candidates from this front, this paper incorporates the computation of knee points along the front. In simpler terms, a knee point denotes a juncture on the Pareto front, where slight improvements in any single objective result in a notable decline in other objectives. Therefore, knee points are of significant interest to decision-makers.

This study utilizes the Kneedle algorithm \cite{51} to estimate knee points. The procedure involves several steps: firstly, all solutions located on the Pareto front are ordered based on one of the objective functions (typically along an axis); subsequently, the sorted solutions are iterated through, and the slope between adjacent solutions is calculated, following the slope calculation described in the introduction of the preceding subsection; finally, the trend of slope values is analyzed to identify a position where the rate of change considerably decelerates, thereby approximating the knee point.

Essentially, by identifying knee points along the front, the process of selecting the most promising candidate AMPs is further refined.

\section{RESULTS}

\subsection{Datasets and Indicators}

\begin{itemize}
\item Datasets: For training samples, sequences with a length of up to 52 amino acids are collected from the following databases: APD\cite{29}, CAMP\cite{30}, LAMP\cite{31}, and DBAASP\cite{32}. After removing redundancy through multiple sequence alignment with a truncation ratio of 0.35, the final datasets comprise 8869 AMP sequences, which is used for HMAMP pretraining. Additionally, 6083 AMPs with experimentally verified low MICs (selected measurement results for E. coli within the range of [-1,3]) from the UniProt database, and 552 hemolytic(HEMO) AMPs from the XGBC-Hem literature\cite{33} focusing on hemolytic activity prediction are chosen to train discriminators. The collected MIC AMP and HEMO data are also employed for transfer learning of AMP attribute classifiers. They are randomly divided into training, validation, and test sets in a ratio of 6:2:2. To demonstrate the generality of our proposed generative model, toxicity datasets from literature\cite{34} are also collected for supplementary experiments. We denote the three dataset distributions as $D_{\text{AMP}}$, $D_{\text{MIC}}$, and $D_{\text{HEMO}}$, where MIC and HEMO correspond to the target attribute datasets. Dataset statistics are presented in Table \ref{tab1}.

\begin{table}[h]
    \caption{Datasets}
    \centering
    \begin{tabular}{|l|c|c|c|}\hline
         \textbf{\textit{Datasets}} &  \textbf{\textit{Training set}}  &  \textbf{\textit{Validation set}} & \textbf{\textit{Test set}} \\
         \cline{1-4}
         $D_{\text{AMP}}$ & 8869 & - & -  \\
         \cline{1-4}
         $D_{\text{HEMO}}$ & 332 & 110 & 110  \\
         \cline{1-4}
         $D_{\text{MIC}}$ & 3651 & 1216 & 1216  \\
         \cline{1-4}
    \end{tabular}
    \label{tab1}
    \centering
\end{table}

\item Hypervolume Indicator: In this article, we employ the hypervolume indicator as a measurement method for finding the balancing efficient set of AMP attributes optimization problems.The concept of hypervolume indicator has been illustated in the PRELIMINARIES section.

\item AMPs Characteristics: We employ the comprehensive analysis methods described in modlAMP\cite{56} to assess various characteristics of AMPs, including attributes like length, charge, hydrophobicity, and hydrophobic moment. In contrast to AMP properties, physicochemical properties like length, charge, hydrophobicity, and hydrophobic moment are easier to compute, but less specific indicators of desired features for peptide sequences.

\end{itemize}

\subsection{Evaluation of HMAMP's Efficacy in Generating Candidate AMPs}
HMAMP represents a pioneering peptide sequence model capable of searching the multi-attribute conditions for AMP activity, encompassing both low MIC and low HEMO. Initially, the teacher mechanism\cite{42} is harnessed to pretrain prior generators using the AMPs dataset ($D_{\text{AMP}}$). Subsequently, the prior generators within the HMAMP framework undergo fine-tuning via two distinct datasets ($D_{\text{MIC}}$ and $D_{\text{HEMO}}$) to generate candidate AMPs tailored for the respective target attributes. In total, 5000 candidate AMPs are generated, constrained by a maximum length of 60, for subsequent experimental scrutiny. The discriminator loss curve of HMAMP is depicted in Figure S3(a). This figure illustrates that the loss for both target discriminators steadily decreases, signifying the stable training and eventual convergence of the HMAMP model.\par

To compare the peptides sequences generated by HMAMP with respect to the three datasets, a comprehensive analysis of their key properties is undertaken. Given the resource-intensive nature of verifying candidate AMPs via wet experiments, we initially focus on properties that are readily measurable and analyzable, such as length, charge, hydrophobicity, and hydrophobic moment. Figure S3(b) showcases that the distribution of the peptides sequences generated by HMAMP closely aligns with the distributions of the datasets. Across most of the 20 natural amino acids, differences are contained within 2\%. Notably, greater disparities are apparent in amino acids like G (glycine), K (lysine), L (leucine), and S (serine). Glycine is hydrophobic, potentially mitigating interactions with cell membranes to reduce hemolysis risk. Conversely, serine is hydrophilic and prone to binding with cell membranes, potentially influencing hemolytic properties. Lysine and leucine, characterized by high positive charge and hydrophilicity, may enhance AMP activity. These subtle amino acid discrepancies affirm the coherence of the generated AMPs with the target attributes. Additionally, charge, hydrophobicity, and hydrophobic moment distributions of candidate AMPs produced by HMAMP are plotted against those of the training set in Figure S3(c)-(f). This reveals a striking resemblance between the attributes of the HMAMP-generated candidates and those of the datasets, attesting to HMAMP's successful assimilation of the potential attribute spaces. Notably, compared to the three datasets, the length distribution of generated AMP concentrates in about 10 and 60, which is consistent with the fact that most AMP with remarkable comprehensive performance are short peptides. In addition, the limit of hyper-parameter length 60 of AMP has a obvious effect on the HMAMP.

Figure S3(b) primarily examines the frequency of individual amino acid occurrences. However, a plethora of studies have elucidated the intricate grammatical structures of peptides\cite{43,44,45,46}. We delve into higher-order amino acid subsequences using word shift methods, assessing the contributions of different subsequences to the sequence, and highlighting their most substantial contributors\cite{47,48}. Figure S4(a)-(c) presents a comparison between HMAMP and subsequences of length 2 across the three datasets. Many of these subsequences are positively charged or hydrophobic, aligning well with the known characteristics of alpha-helical AMPs. Notably, in HMAMP's subsequence contribution diagram, sg, gp, and yg stand out prominently. The presence of these subsequences exerts a direct influence on the interaction between AMPs' secondary structures and their targets, ultimately affecting AMP activity and hemolysis.

When generating candidate AMPs, it is essential that the generated peptides exhibit diversity and novelty in comparison to known AMPs. If the sequence diversity produced by HMAMP is low, it might indicate the occurrence of the modal collapse issue mentioned earlier. In this study, the Gotoh global alignment algorithm is employed to quantify sequence diversity\cite{49,50}, where a higher alignment score between two sequences signifies a higher degree of similarity. Figure S5(a) presents the overall score graph illustrating the diversity of the peptides generated by HMAMP compared to the datasets. It is evident that those generated by HMAMP exhibits higher median, mean, lower median, and upper percentile scores than the three datasets ($D_{\text{AMP}}$, $D_{\text{MIC}}$, and $D_{\text{HEMO}}$), respectively, indicating a higher diversity value.

\begin{figure}
\centering 
\includegraphics[width=1\columnwidth]{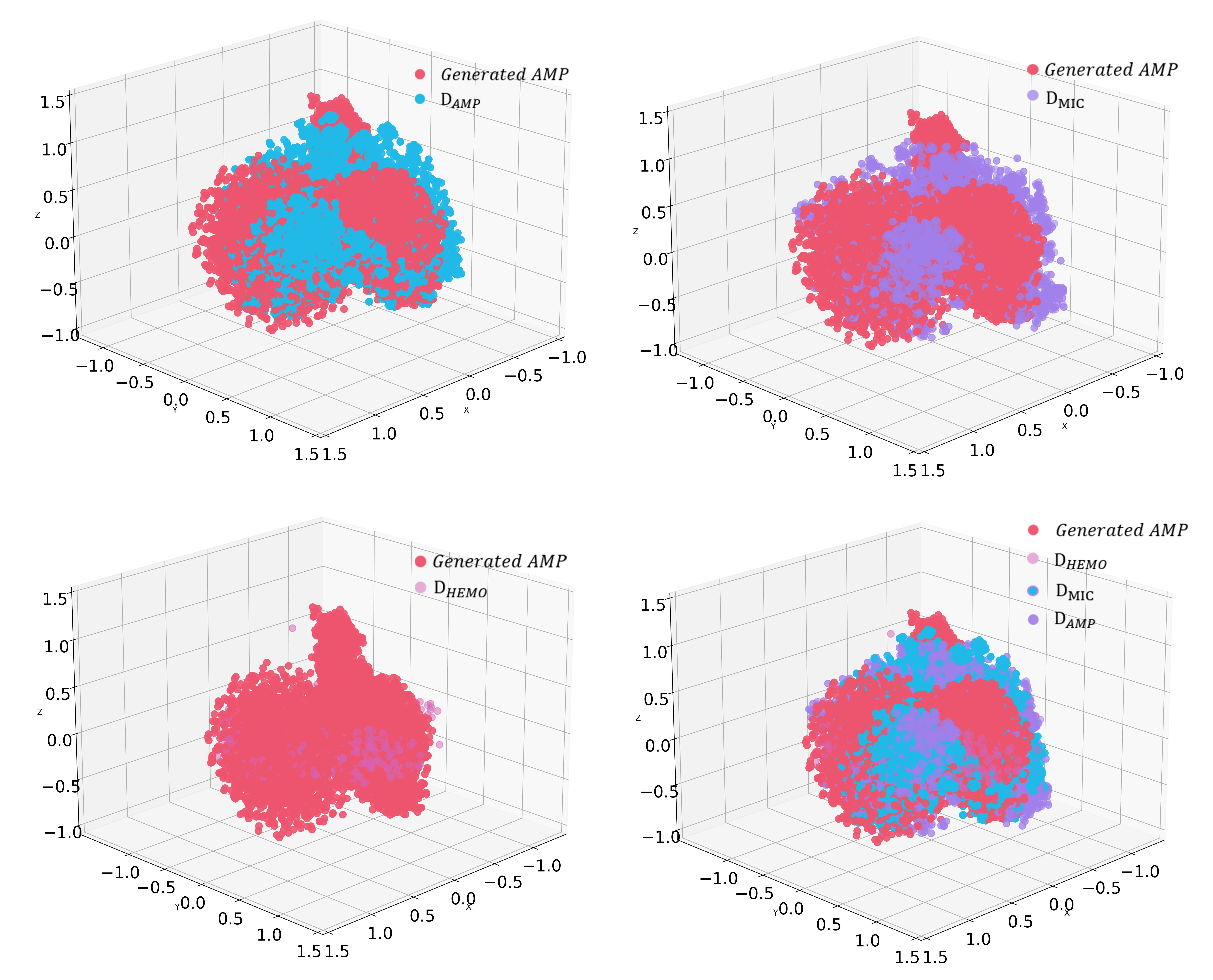}
\caption{The t-SNE plot illustrating the distribution of amino acids in each sequence generated by HMAMP and three datasets ($D_{\text{AMP}}$, $D_{\text{MIC}}$, and $D_{\text{HEMO}}$) .}
\label{visualization}
\end{figure}

To provide further insight into the spatial distribution of candidate AMPs generated by HMAMP, we present the t-SNE distribution of these AMPs in comparison to the three datasets in Figure \ref{visualization}. We employ the PCA algorithm\cite{57} to transform high-dimensional sequence feature vectors into a 3D space for visualization. As shown in Figure \ref{visualization}, the spatial distribution of sequences generated by HMAMP exhibits a significant overlap with the distributions of the three datasets, indicating a substantial degree of coverage. This observation suggests that HMAMP captures the global distribution of the datasets and has the ability to explore a wider potential sequence space.

\subsection{Comparing HMAMP with State of the Arts}

We conducted a comparative evaluation with five benchmark methods, spanning language models, encoder and adversarial generative networks, and evolutionary algorithms. We replicated the following methods in our experiments:

\begin{itemize}
    \item \textbf{Generative LSTM by Muller et al.\cite{37}}: This method utilizes a generative LSTM designed for de novo peptides. It was trained on a dataset consisting of 1554 AMPs extracted from the ADAM\cite{58}, APD\cite{29}, and DADP\cite{59} databases to generate active peptides.
    
    \item \textbf{AMP-GAN by Tzu et al.\cite{38}}: AMP-GAN is a deep convolutional neural generative adversarial network that generates candidate active AMPs with a broad spectrum. It was trained to generate AMPs with desired properties.
    
    \item \textbf{PepGAN by Zhang et al.\cite{39}}: PepGAN is an adversarial model tailored for AMPs, aimed at generating high-fidelity and highly active peptides. It focuses on generating AMPs with specific attributes.
    
    \item \textbf{Wasserstein Autoencoder (WAE) by Yang et al.\cite{40}}: Yang et al. combined the Wasserstein autoencoder with a particle swarm optimization forward search algorithm to screen anticancer peptides with desired attributes. We adapted this approach for AMP datasets to assess the generation performance of the encoder.
    
    \item \textbf{AMPEMO by Liu et al.\cite{41}}: AMPEMO is a multi-objective evolutionary method designed to optimize both the antimicrobial activity and diversity of AMPs. Liu et al. proposed this method for generating AMPs with enhanced attributes.
\end{itemize}

These benchmark methods were evaluated alongside HMAMP to comprehensively assess their performance in generating AMP candidates with desired attributes.

We conducted an analysis of the fundamental properties of the 5000 candidate AMPs generated by each benchmark methods, comparing them with HMAMP. Figure S6(a)-(e) presents a visualization of key properties including the distribution of amino acids, length, charge (C), hydrophobicity (H), and hydrophobic moment (HM) for both the HMAMP method and the other baseline methods. For each property, we specified a range: charge ($2 <$ C $< 5$), hydrophobicity (H$ > 0.25$), and hydrophobic moment ($0.5 <$ HM $< 0.75$, HM $> 1.75$). We measured the percentage of peptides within each property range to assess the generation performance. Additionally, we computed the combination of these percentages to measure the proportion of peptides satisfying all three criteria simultaneously. 

\begin{figure*}
\centering 
\includegraphics[width=\textwidth]{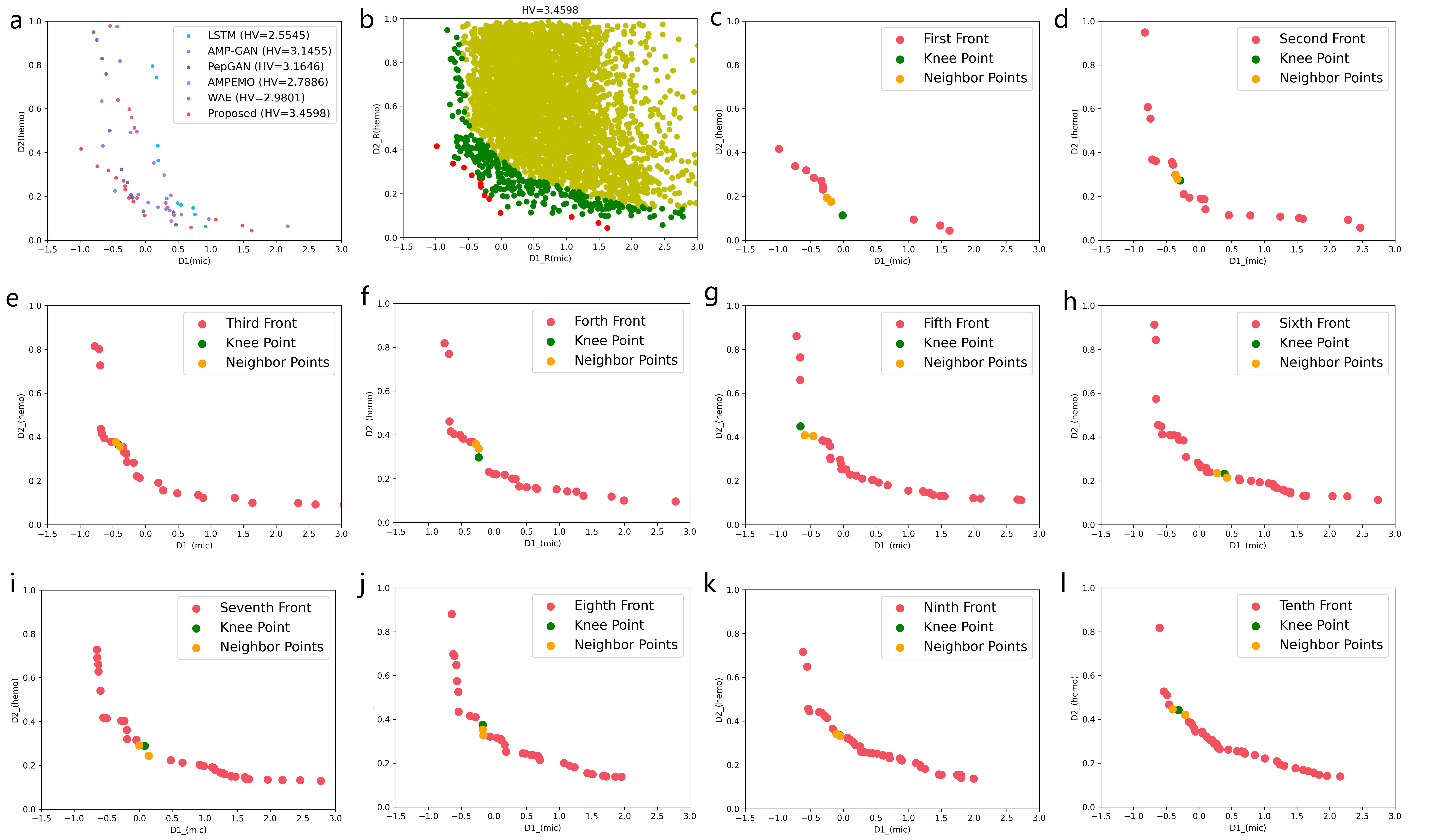}
\caption{a. HV performance comparison between HMAMP and benchmark methods for MIC and hemolytic attributes. b. Distribution of AMPs generated by HMAMP for MIC and hemolysis targets. c-l. Top ten Pareto fronts for hemolytic and antimicrobial activity, with orange points indicating knee point solutions.}
\label{pareto set}
\end{figure*}

For more intuitive comparison, Table \ref{tab2} summarizes the performance of the generation process, showing the percentage of peptides falling within each attribute range, as well as the percentage of candidate AMPs meeting all three attribute criteria. Moreover, we employed trained classifiers to predict hemolysis and MIC values for the 5000 candidate AMPs. These predicted values were then used as the two objectives for multi-objective optimization, and the hypervolume (HV) metric was used to comprehensively evaluate the generation performance of HMAMP and the five benchmark methods. The results in Table \ref{tab2} reveal that the proportion of candidate AMPs falling within the proposed method's combination interval is higher than that of the other benchmark methods, except for WAE. In terms of HV value, HMAMP significantly outperforms the other comparison methods, indicating that the convergence and distribution of the solution set of the 5000 candidate AMPs generated by HMAMP are super. These results are visually represented in Figure \ref{pareto set}(b).

\begin{table}
    \caption{Comparative Experiments of HMAMP and Benchmark Methods.}
    \centering
    \begin{tabular}{|c|c|c|c|c|c|c|}\hline
         \textbf{\textit{ }} &  
         \textbf{\textit{C }} &
         \textbf{\textit{H}}  &  
         \textbf{\textit{HM}} &
         \textbf{\textit{Combination}} &
         \textbf{\textit{HV}} \\
         \cline{1-6}
         LSTM & 0.0322 & 0.1269 & 0.0915 & 0.0915 & 2.5545  \\
         \cline{1-6}
         AMP-GAN & 0.9176 & 0.15 & 0.0142 & 0.0142 & 3.1455 \\
         \cline{1-6}
         PepGAN & 0.498 & 0.2252 & 0.074 & 0.074 & 3.1646  \\
         \cline{1-6}
         AMPEMO & 0.2384 & 0.3665 & 0.0896 & 0.0896 & 2.7886 \\
         \cline{1-6}
         WAE  & 0.5922 & 0.3282 & 0.1849 &0.1849 &2.9801  \\
         \cline{1-6}
         Proposed  & 0.4990 & 0.2330 & 0.0958 & 0.0958 & 3.4598  \\
         \cline{1-6}
    \end{tabular}
    \label{tab2}
    \centering
\end{table}

In addition, we conducted diversity comparison experiments between HMAMP and the five benchmark methods. Similar to the previous comparison experiments, we employed the Gotoh global alignment algorithm to quantitatively assess the diversity of sequences. Looking at Figure S5(b), we observe that, except for the LSTM benchmark method, the median and upper percentiles of diversity scores for the other benchmark methods are greater than those of HMAMP. This indicates that among the benchmark methods, HMAMP stands out as the most promising solutions for generating diverse and novel candidate AMPs.

Due to the impracticality of analyzing all 5000 candidate AMPs individually, we focused on the solution sets on the top 5 Pareto fronts. Utilizing the distance-based knee point method, we identified several knee points that satisfy multi-objective attributes, as depicted in Figure \ref{pareto set}(c)-(l). Intuitively, neighboring points near knee points tend to be significant potential AMP. We attached orange to the neighborhood points in Figure \ref{pareto set}(c)-(l), \ref{pareto set toxi}(c)-(l) did the same. The discussion of these ideal AMP candidates will be further elaborated in Part D of the RESULTS.

\subsection{Additional Analysis: Multi-objective AMPs Generation for MIC and Toxicity}

\begin{figure*}[htbp]
\centering 
\includegraphics[width=\textwidth]{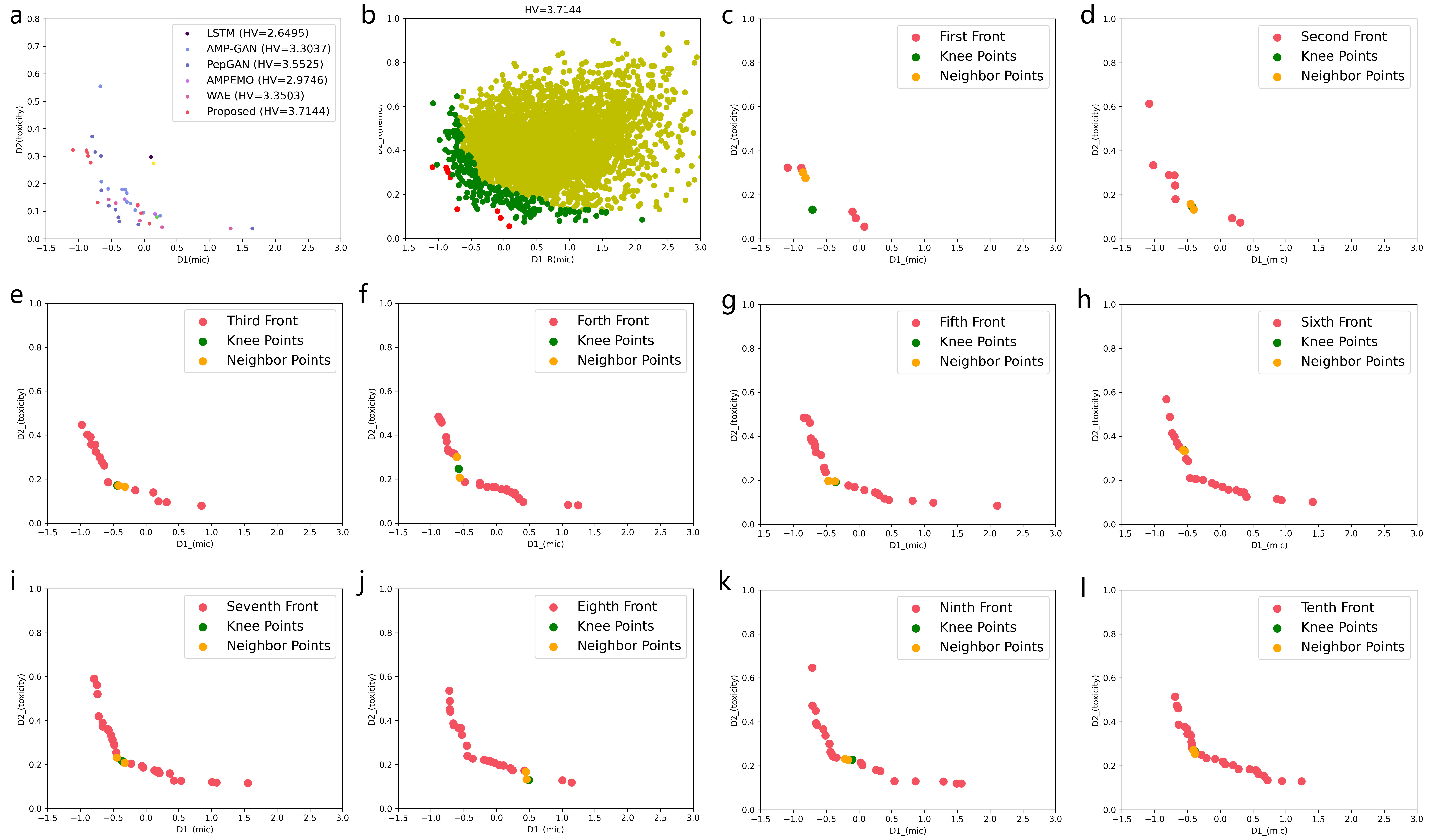}
\caption{a. The Pareto front and HV Performance comparison between HMAMP and benchmark methods for MIC and toxicity. b. Distribution of AMPs generated by HMAMP for MIC and toxicity targets. c-l. The  first 10 non-dominated fronts, respectively, with the orange points representing knee point solutions.}
\label{pareto set toxi}
\end{figure*}

To further showcase the effectiveness and general applicability of our proposed HMAMP framework in generating multi-objective attribute candidate AMPs, we conducted additional experiments involving both MIC and toxicity attributes. In this case, we trained a multi-objective generation network using existing toxicity datasets and fine-tuned the Prot-BERT-BFD model to create an efficient toxicity predictor. Similarly, we compared the results with the same five benchmark methods as previously mentioned. The experimental outcomes are depicted in Figure \ref{pareto set toxi}(a)-(l).

We acquired the solution set by predicting AMPs using both the MIC predictor and the toxicity classifier. While some methods, including the first non-dominated front solution set of our proposed HMAMP, exhibit fewer solutions, and even certain duplicate solutions appear, no distinct advantages can be observed from the solutions in the first Pareto front. However, the overall HV value of our proposed HMAMP method significantly surpasses that of other methods. This indicates that the distribution and convergence of the solution set offered by HMAMP are notably excellent.

Furthermore, we observed that the solution set in the upper-left region is relatively sparse. This scarcity might be attributed to the limited availability of experimentally confirmed toxic AMPs. To identify the AMPs that best balance antimicrobial peptide activity and toxicity, we calculated 5,000 candidate AMPs to explore the first 10 non-dominated fronts. Subsequently, we identified the most ideal knee point, representing the most promising candidate AMPs. The detailed analysis of these AMP candidates will be presented in Subsection V.E.

\subsection{Analysis and Structural Prediction of Potential AMPs}

The majority of discovered effective AMPs, although diverse in sequence, structure, and length, share two common attributes: cationic nature and active amphiphilicity. Most cationic AMPs are comprised of 12-60 amino acid residues, having an isoelectric point range of 8.9-12, a charge range of 2-7, and a hydrophobic moment range of 0.2-1.0, which imparts strong cationic characteristics. Under specific conditions, the N-terminal of AMPs can readily adopt an $\alpha$-helical structure, endowing them with amphiphilic properties that facilitate interaction with the membrane structures of target cells or pathogens. This $\alpha$-helical conformation is also a crucial structural feature for AMPs in exerting their antimicrobial activity.

Based on these common characteristics of AMPs, we calculated the knee points from the first 10 non-dominated fronts and ultimately selected the 10 most intriguing candidate AMPs for two sets of conflicting targets.

Table \ref{tab3} presents the results of molecular weight (M), isoelectric point (E), charge quantity (C), hydrophobic moment (HM), MIC values, hemolysis probability, and toxicity probability for these 10 candidate AMPs. IDs 1 to 5 represent candidates generated with MIC and hemolysis targets. It's notable that the highest and lowest MIC values among these candidates are 1.592 $\upmu$g/mL and 0.592 $\upmu$g/mL, respectively, indicating significant antibacterial potency. The highest predicted hemolysis probability is only 0.388, suggesting a relatively low likelihood of hemolytic properties for these candidates. IDs 6 to 10 represent candidates generated with MIC and toxicity targets. Among these, the highest and lowest MIC values are 2.075 $\upmu$g/mL and 0.455 $\upmu$g/mL, respectively. The highest predicted toxicity probability is 0.375, indicating that the multi-objective optimization process has yielded AMPs with high activity and low toxicity. Additionally, the isoelectric point, charge quantity, and hydrophobic moment ranges of these candidates are within reasonable bounds.

For the MIC and hemolysis targets, we employed the Alphafold\cite{54} method to predict their three-dimensional structures. Figure S7(a)-(e) and Figure  S8(a)-(e) display the Predicted Alignment Error map (PAE) and the Prediction Interface Distance Difference Test (PIDDT) plot for the generated candidate sequences ID1-ID5. The PAE plot visualizes the differences between the positions of residues in the predicted structure and their positions in the actual structure. Lower color intensities indicate smaller errors, reflecting higher prediction accuracy. On the other hand, higher color intensities, particularly in red, suggest less reliable predictions. The PIDDT plot showcases the differences between the predicted and actual distances between residues. A high PIDDT value indicates close agreement between the predicted and actual distances, implying higher accuracy. In contrast, lower PIDDT values point to significant deviations between predicted and actual distances. Values closer to 1 signify more accurate predictions. We observed that IDs 1 to 5 exhibit lower color intensities and higher PIDDT values, indicating that these models produce more accurate predictions for these candidates. Notably, ID4 displayed suboptimal PIDDT predictions at position 15, possibly due to Alphafold's preference for analyzing shorter peptides.

For the MIC and toxicity targets, the results of the Alphafold predictive analysis are shown in Figure S9(a)-(e) and S10(a)-(e). Here, the lower color intensities and higher PIDDT values for IDs 6 to 10 suggest that these models' predictions are generally more accurate for this set of candidates. These predictions are slightly inferior to the overall prediction results for IDs 1 to 5, possibly due to the scarcity of toxicity experimental data or the greater flexibility of protein structures.

We also visually depicted the spiral wheel diagrams and secondary structures of the candidate sequences ID1-ID5 and ID6-ID10, as shown in Figure S11(a)-(b) and S12(a)-(b), respectively. We observed that these candidate AMPs are rich in positively charged amino acid side chains (arginine R and lysine K), enabling them to interact with negatively charged components on bacterial surfaces, thereby enhancing their antibacterial activity. Additionally, they contain numerous hydrophilic amino acid side chains (aspartic acid N, glutamine Q, glutamate E, serine S, threonine T, and tyrosine Y) that facilitate interaction with ions in water and solutions. This interaction helps mitigate hemolysis, as it enhances the stability of AMPs in solution, reducing their direct contact with red blood cell membranes. For the MIC and toxicity targets, these candidate AMPs possess abundant hydrophilic amino acids (serine S, threonine T, and aspartate N), which may modulate their conformation and toxicity by interacting with hydrophobic regions. Additionally, their relatively low hydrophobic moments decrease the likelihood of binding to host cells and reduce potential host cell toxicity. Figure S11(b) and S12(b) indicate that these candidate AMPs predominantly adopt an $\alpha$-helical conformation, the most common structural class of AMPs. While ID3, ID5, and ID8 exhibit flaky secondary structures, these flaky AMPs may exhibit varying activities against different pathogens and in diverse application scenarios. Due to their relatively simple structures, they generally demonstrate high stability and are suitable candidates for potential anti-infection drugs.

\begin{table*}[h]
\caption{Top Candidate AMPs}
\centering
\begin{tabular}{|c|c|c|c|c|c|c|c|c|}
\hline
ID & Sequence                           & M        & E      & C     & HM    & MIC Value & Hemolysis Probability & Toxicity Probability \\ \hline
1  & rkvfrrvvp                          & 1156.427 & 11.999 & 3.555 & 0.849 & 1.211     & 0.163                 & ----                 \\ \hline
2  & iriailplarkwmvws                   & 1953.442 & 11.999 & 2.555 & 0.288 & 1.293     & 0.36                  & ----                 \\ \hline
3  & ilkyckrtv                          & 1123.412 & 9.787  & 2.525 & 0.755 & 1.592     & 0.362                 & ----                 \\ \hline
4  & arrykffirlrqrlvknygrvitgletyrdligp & 4169.88  & 11.327 & 6.597 & 0.356 & 0.592     & 0.388                 & ----                 \\ \hline
5  & ckdwavknyykiykg                    & 1879.185 & 9.364  & 2.516 & 0.356 & 1.531     & 0.265                 & ----                 \\ \hline
6  & kwfvvwisklvsklsnnp                 & 2145.544 & 10.302 & 2.55  & 0.412 & 1.149     & ------               & 0.259                \\ \hline
7  & llplllkfflskstv                    & 1719.158 & 10.003 & 1.552 & 0.199 & 2.075     & ------               & 0.28                 \\ \hline
8  & illpalgllipsiscsmnkrcnlhlgt        & 2878.522 & 8.957  & 1.543 & 0.092 & 1.465     & ------               & 0.335                \\ \hline
9  & vlwilflgfwilltkkmsekfrrrly         & 3358.138 & 11.069 & 4.514 & 0.215 & 1.647     & ------               & 0.375                \\ \hline
10 & kifmqiltkikktaknvsetiqtnky         & 3069.66  & 10.126 & 4.541 & 0.462 & 0.455     & ------               & 0.283                \\ \hline
\end{tabular}
\label{tab3}
\end{table*}

\section{Conclusion}

In this study, we introduced a novel approach, Hypervolume-driven Multi-objective Antimicrobial Peptide Generation (HMAMP), for the de novo design of antimicrobial peptides (AMPs) with desired attributes.  HMAMP takes a hybrid approach by integrating generative adversarial and reinforcement learning techniques. It employs the concept of Hypervolume maximization through gradient descent for training. Leveraging the power of multi-objective optimization to simultaneously optimize multiple conflicting attributes, such as antimicrobial activity and hemolysis, HMAMP achieves a balance between the desired attributes, leading to the generation of diverse and highly potent AMP candidates. Based on the Pro-BERT-BFD model, we meticulously fine-tune two predictive models to estimate attribute values for the generated candidate AMPs. This enables the acquisition of the Pareto front and knee points of the candidate AMPs.

Our extensive experiments demonstrated the effectiveness and superiority of HMAMP in generating candidate AMPs with desired properties. By comparing HMAMP with five benchmark methods, including generative LSTM, deep convolutional neural networks, and evolutionary algorithms, we showcased HMAMP's ability to generate AMPs with higher diversity and potency. The results obtained through diverse evaluation metrics, such as Gotoh alignment scores and Hypervolume (HV), confirmed that HMAMP consistently outperforms other methods.

Moreover, we showcased HMAMP's versatility by performing additional  multi-attribute AMP generation experiments. These experiments underscored HMAMP's ability to effectively generate AMPs with different sets of attributes, including antimicrobial activity and toxicity, further highlighting its broad applicability.

Through comprehensive structural and attribute analysis, we identified a selection of top candidate AMPs generated by HMAMP. These candidates exhibited characteristics in line with established patterns of effective AMPs, including cationic nature, amphiphilicity, and favorable structural features. Furthermore, the Alphafold predictions of these candidate AMPs' secondary structures provided valuable insights into their potential mechanisms of action and interactions with target membranes.

In summary, our proposed HMAMP framework presents a powerful and versatile solution for the design of antimicrobial peptides with desired attributes. By addressing the multi-objective nature of AMP design, HMAMP demonstrates the potential to revolutionize the field of antimicrobial peptide development, leading to the discovery of new and effective therapeutic agents for combating microbial infections and contributing to the broader field of drug discovery.

Future work could involve refining the multi-objective optimization process, leveraging additional attribute predictors, and exploring experimental validations to confirm the antimicrobial and therapeutic potential of the generated candidates. The integration of advanced molecular dynamics simulations and in vitro studies could further enhance our understanding of the candidates' mechanisms of action and pave the way for their eventual translation into practical applications.

\bibliographystyle{IEEEtran}
\small\bibliography{HMAMP}

\end{document}


\title{HMAMP: Hypervolume-Driven Multi-Objective Antimicrobial Peptides Design

\{Supplementary Material\}}

\onecolumn

\author{Li~Wang, Yiping~Liu~\Letter, Xiangzheng Fu~\Letter, Xiucai~Ye, Junfeng Shi, Gary G. Yen, Fellow, IEEE, Xiangxiang Zeng
\thanks{Corresponding Authors: Yiping Liu (yiping0liu@gmail.com), Xiangzheng Fu(excelsior511@126.com); Other Authors:Li~Wang(wl12345678@hnu.edu.cn), Xiucai~Ye(yexiucai@cs.tsukuba.ac.jp), Junfeng~Shi(Jeff-shi@hnu.edu.cn), Gary G. Yen(gyen@okstate.edu), Xiangxiang Zeng 
(xzeng@foxmail.com) } 
\thanks{This work was supported by the National Natural Science Foundation of China (Grant No. 662002111,  62372158, 62106073, 62122025, U22A2037, 62250028), the Natural Science Foundation of Hunan Province (Grant No. 2022JJ40090)} 

\thanks{
Li~Wang, Yiping~Liu, X. Zeng, and Junfeng Shi are with the College of Computer Scienceand Electronic Engineering, Hunan University, Changsha, Hunan,
(wl12345678@hnu.edu.cn, yiping0liu@gmail.com, xzeng@foxmail.com, Jeff-shi@hnu.edu.cn)

Gary G. Yen is with the School of Electrical and Computer Engineering, Oklahoma State University, Stillwater. OK 74078. USA. (gyen @okstate.edu)

Xiucai~Ye is with the School of System Information and Engineering, University of Tsukuba,Ibaraki, Japan(yexiucai@cs.tsukuba.ac.jp)

Xiangzheng Fu is with Neher's Biophysics Laboratory for Innovative Drug Discovery, State Key Laboratory of Quality Research in Chinese Medicine, Macau Institute for Applied Research in Medicine and Health, Macau University of Science and Technology ,Macau, China, 999078 (e-mail:excelsior511@126.com).
}
}

\markboth{  }%
{Shell \MakeLowercase{\textit{et al.}}: Bare Demo of IEEEtran.cls for IEEE Journals}

\maketitle

This additional material mainly presents definitions related to multi-objective optimization and GANs; the loss curves of discriminators; basic property distributions for three datasets ($D_{\text{AMP}}$, $D_{\text{MIC}}$, and $D_{\text{HEMO}}$) and AMPs generated by HMAMP; shannon’s entropy divergence; Letter-value plots;  basic property distributions of AMPs generated by five benchmark
methods and AMPs generated by HMAMP; Predicted Alignment Error maps (PAE); Prediction Interface Distance Difference Test plots.

\section{Multi-objective Optimization}

A multi-objective optimization problem is commonly defined as follows:

\begin{equation}
\underset{x \in \Omega}{\min} F(x) = \left[\begin{array}{c} f_1(x) \\ f_2(x) \\ \vdots \\ f_n(x) \end{array}\right],
\end{equation}

where $\Omega$ represents a discrete search space, and $F(x)$ is an $n$-dimensional objective vector. Due to conflicting objectives, no single solution can simultaneously optimize all objectives. Instead, a collection of Pareto optimal solutions exists, as defined below.

\textit{Definition 1} (Pareto Dominance): For $x_1, x_2 \in \Omega$, $x_1$ is said to dominate $x_2$ (denoted as $x_1 \prec x_2$) if and only if $f_i(x_1) \leq f_i(x_2)$ for all $i \in \{1, ..., n\}$ and $f_j(x_1) < f_j(x_2)$ for some $j \in \{1, ..., n\}$.

\textit{Definition 2} (Pareto Optimal Solution): A solution $x^* \in \Omega$ is a Pareto optimal solution if no $\hat{x} \in \Omega$ exists such that $\hat{x} \prec x^*$. The collection of all Pareto optimal solutions is referred to as the Pareto set, while the map of the Pareto set in the objective space is termed the Pareto front.

In this paper, we treat the task of AMP design as a multi-objective optimization problem, where the objectives encompass the optimization of various attributes of AMPs, such as elevated activity and diminished toxicity.

\section{Knee Points}

\begin{figure}
    \centering
    \includegraphics[width=0.8\linewidth]{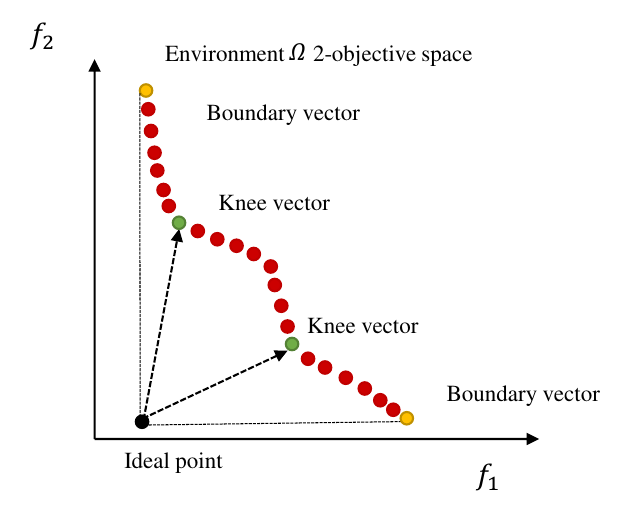}
    \caption{Knee points in the Pareto front.}
    \label{knee_point}
\end{figure}

Knee points constitute a subset of Pareto optimal solutions distinguished by the characteristic that a marginal enhancement in one objective results in a significant deterioration of the others\cite{72,73}. This phenomenon discourages movement in any direction, making knee points preferable locally in the absence of specific user preferences. In this study, akin to \cite{53}, we employ the mathematical definition of curvature for a continuous function as the foundation of our knee point definition. The curvature is computed as follows:

\begin{equation}
\kappa = \frac{|x'y'' - y'x''|}{(x'^2 + y'^2)^{3/2}},
\end{equation}

where $x'$ and $y'$ denote the first derivative of the curve at that point, while $x''$ and $y''$ symbolize the second derivative of the curve at that point. The value $\kappa$ signifies the curvature at that particular point. More specifically, the subsequent formula can be employed to calculate the curvature at the $i$-th point:

\begin{equation}
\kappa_i = \frac{|(x_i - x_{i-1})(y_{i+1} - y_i) - (y_i - y_{i-1})(x_{i+1} - x_i)|}{((x_i - x_{i-1})^2 + (y_i - y_{i-1})^2)^{3/2}},
\end{equation}

where $x_i$ and $x_j$ represent the coordinates of the $i$-th point on the Pareto front, and $\kappa_i$ denotes the curvature of that point. For instance, in Figure \ref{knee_point}, the green dot exhibits the most pronounced inflection relative to both boundary points, making it the designated knee point of the current solution.

\section{Hypervolume Indicator}

\begin{figure}
\centering 
\includegraphics[width=1\columnwidth]{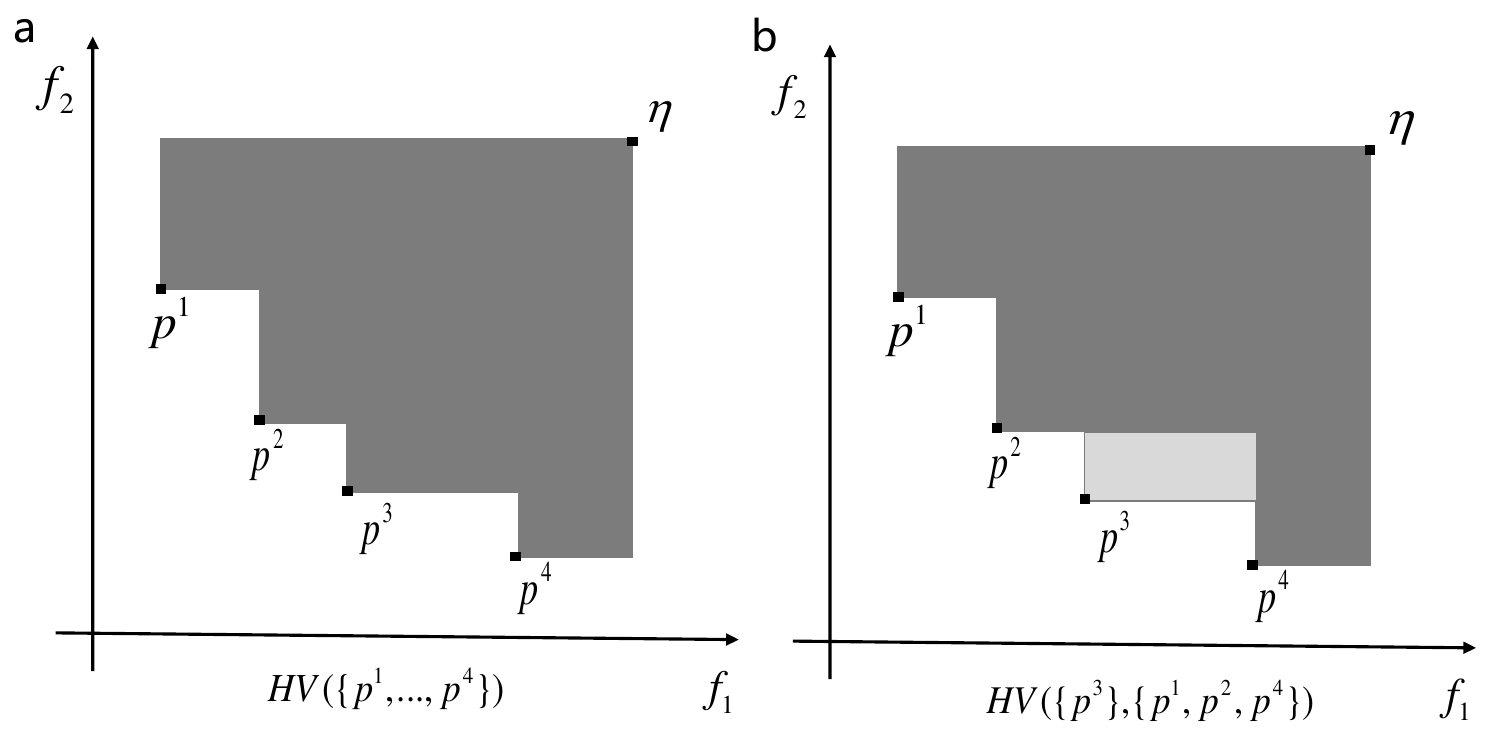}
\caption{Example of hypervolume for a two-objective minimization problem.}
\label{Hypervolume}
\end{figure}

The Hypervolume (HV) indicator was initially introduced as a means of evaluating a solution set in multi-objective optimization\cite{35,36}. It assesses the results of an optimizer by considering the convergence, diversity, and distribution of points with respect to the Pareto front. Given a set of Pareto optimal front solutions $S\subset \mathbb{R}^{k}$ and a reference point $\eta^*$, the HV indicator of $S$ quantifies the volume of the region that is weakly dominated by $S$ and bounded above by $\eta^*$:
\begin{equation}
HV(\mathrm S)=\Lambda\left(\bigcup\limits_{p\in S, p\leq\eta^*}[p,\eta^*]\right),
\end{equation}
where $\Lambda(\cdot)$ represents the Lebesgue measure. Figure \ref{Hypervolume}(a) illustrates a two-dimensional instance of the HV concept. The choice of $\eta^*$ can directly impact the computed HV value. 
The HV contribution of a point set to some reference point set\cite{69,70} is formally
defined based on the definition of HV indicator, the HV contribution of point $p$ to point set $S$ is:
\begin{equation}
HV(p,S)=HV(S\cup\{p\})-HV(S\setminus\{p\}).
\end{equation}
The hypervolume contribution of a point is sometimes referred in the literature as the incremental
hypervolume or the exclusive hypervolume\cite{71}. Moreover, the contribution of a point p to the empty set is sometimes called the inclusive hypervolume \cite{71}. See Figure\ref{Hypervolume}(b) for two-dimensional example of a hypervolume contribution.

As we are dealing with a multi-objective minimization problem, the reference point $\eta^*$ is positioned in the upper right corner. A larger HV value signifies a well-converged and well-distributed solution set.

\section{Generative Adversarial Networks}\label{supplement:S1}

In brief, the Generative Adversarial Network (GAN) model consists of two pivotal components: a generative network module G and a discriminator network module D \cite{25}. The role of G is to learn a sample distribution similar to $P_z$ given an actual data distribution, while D takes points within the data space and computes the probability that $x$ is drawn from $P_{data}$ rather than being generated by G. Guided by D, G is trained to produce data that maximizes D's probability, until D becomes incapable of distinguishing between generated and authentic data. These two networks are alternatively trained in a minimax adversarial game:

\begin{equation}
\min_G \max_D \mathbb{E}_{\mathbf{x}\sim P_{data}}\left[\log\left(D\left(\mathbf{x}\right)\right)\right] + \mathbb{E}_{\mathbf{z}\sim P_{\mathbf{z}}}\left[\log\left(1-D\left(G\left(\mathbf{z}\right)\right)\right)\right],
\end{equation}

where $P_{data}$ signifies the data distribution, $x$ denotes a point within the data space, $z$ represents samples from the latent distribution, and $P_{\mathbf{z}}$ corresponds to the model distribution.
\clearpage 

\onecolumn

\begin{figure*}[htbp]
\centering 
\includegraphics[width=\textwidth]{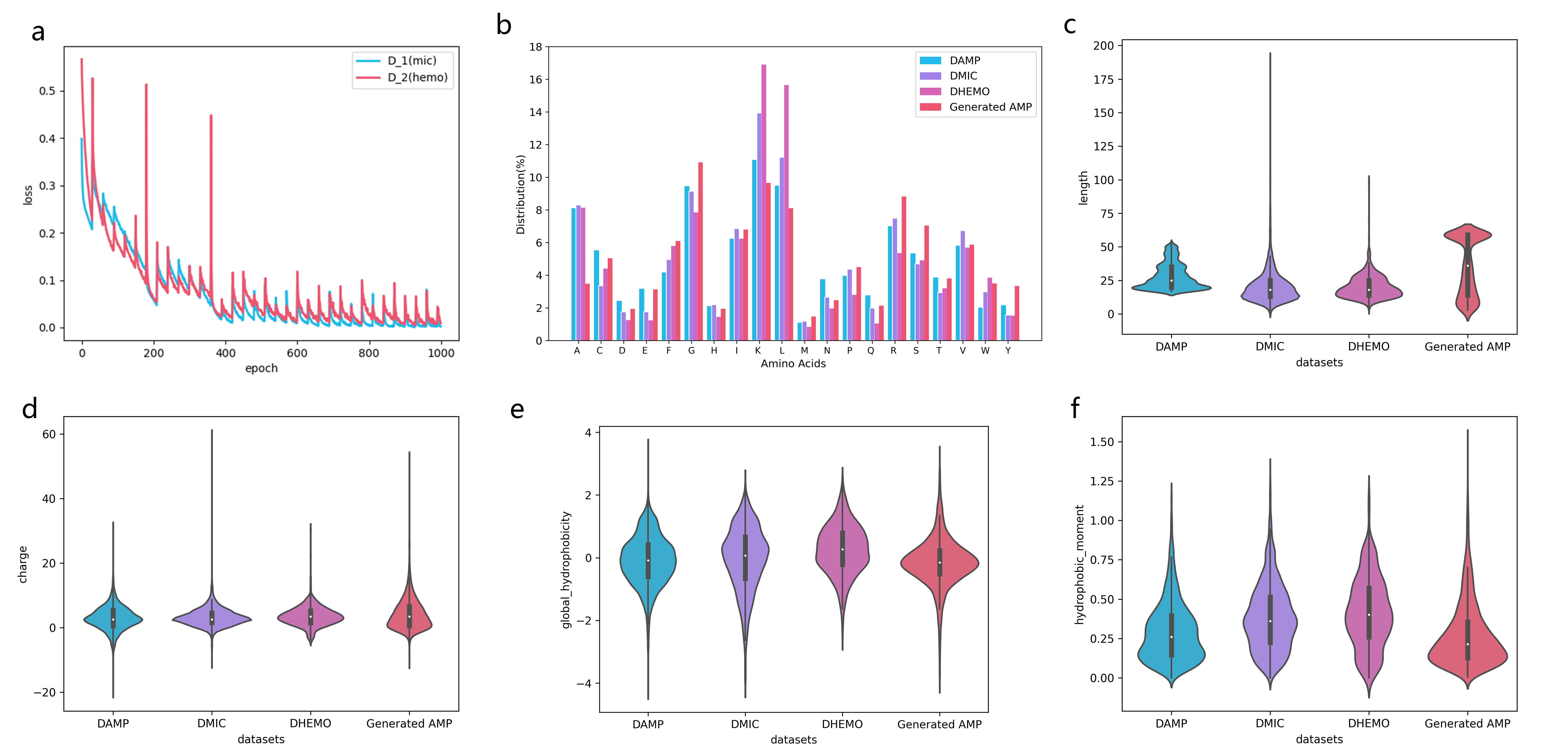}
\caption{a. The loss curves of Discriminators. b-f. Distributions of amino acids, sequence length, charge, hydrophobicity, and hydrophobic moment for three datasets ($D_{\text{AMP}}$, $D_{\text{MIC}}$, and $D_{\text{HEMO}}$) and AMPs generated by HMAMP, respectively.}
\label{S1}
\end{figure*}

\clearpage 
\begin{figure*}[htbp]
\centering 
\includegraphics[width=\textwidth]{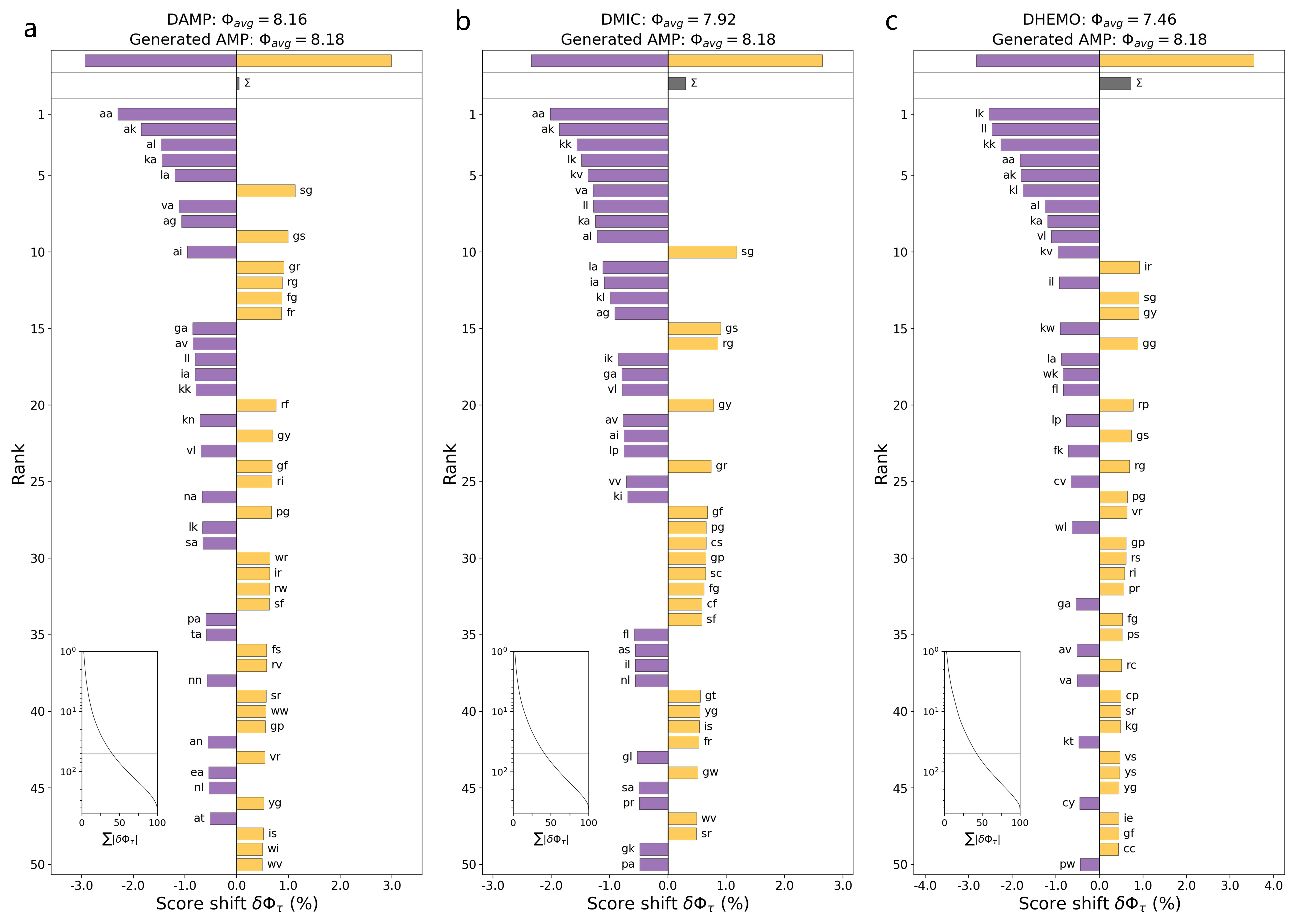}
\label{fig6}
\caption{Shannon's entropy divergence between the distributions of length 2 subsequences of FASTA characters in AMPs from those generated by HMAMP and the datasets ($D_{\text{AMP}}$, $D_{\text{MIC}}$, and $D_{\text{HEMO}}$). Purple bars indicate a higher prevalence of a specific subsequence in real AMPs (i.e., AMPs in the datasets), while gold bars indicate a higher prevalence in generated AMPs. The two values in the title of each panel represent the average entropy of each group. The cumulative distribution function (CDF) plot in the lower left corner of each panel illustrates that the top 50 contributors to the divergence only encompass 50\%, emphasizing the notably flat nature of all distributions.}
\label{S2}
\end{figure*}
\clearpage 

\begin{figure*}[htbp]
\centering 
\includegraphics[width=\textwidth]{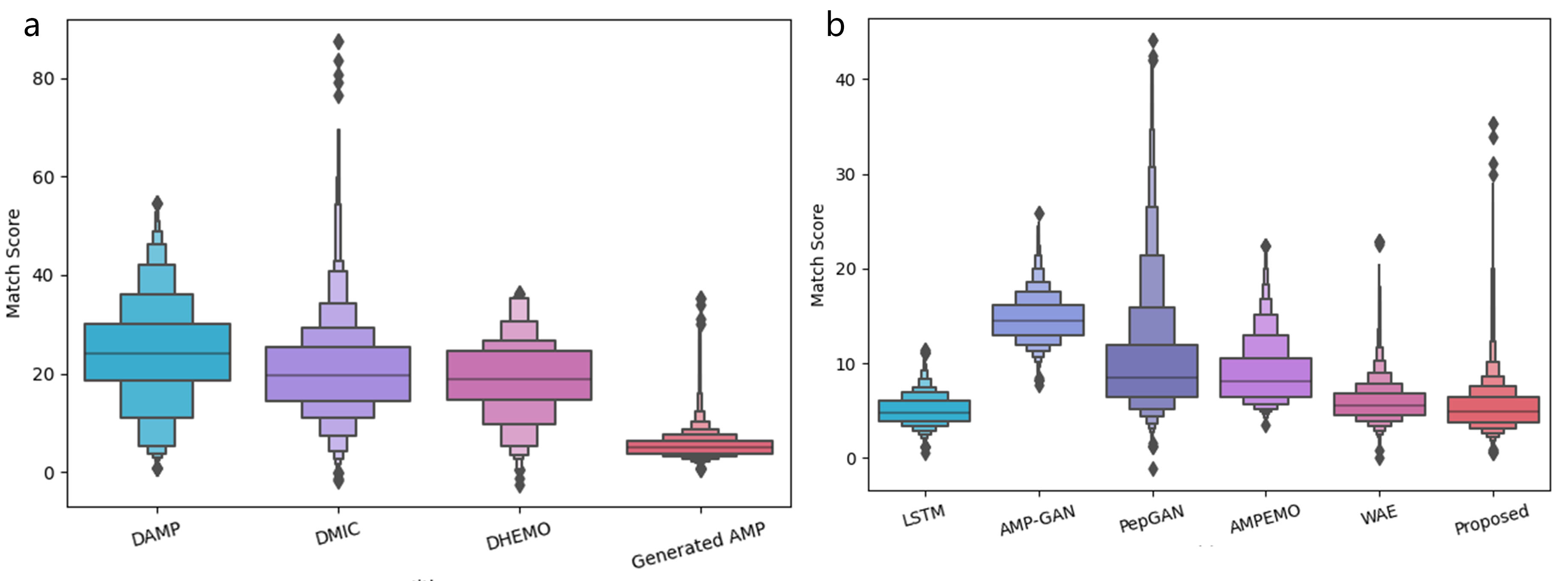}
\caption{a.Letter-value plots showing distributions of match scores comparisons between HMAMP(generated AMP) and  three datasets. b. Letter-value plots showing distributions of match scores comparisons between  HMAMP(proposed) and  five benchmark methods.}
\label{S3}
\end{figure*}
\clearpage 

\begin{figure*}[htbp]
\centering 
\includegraphics[width=\textwidth]{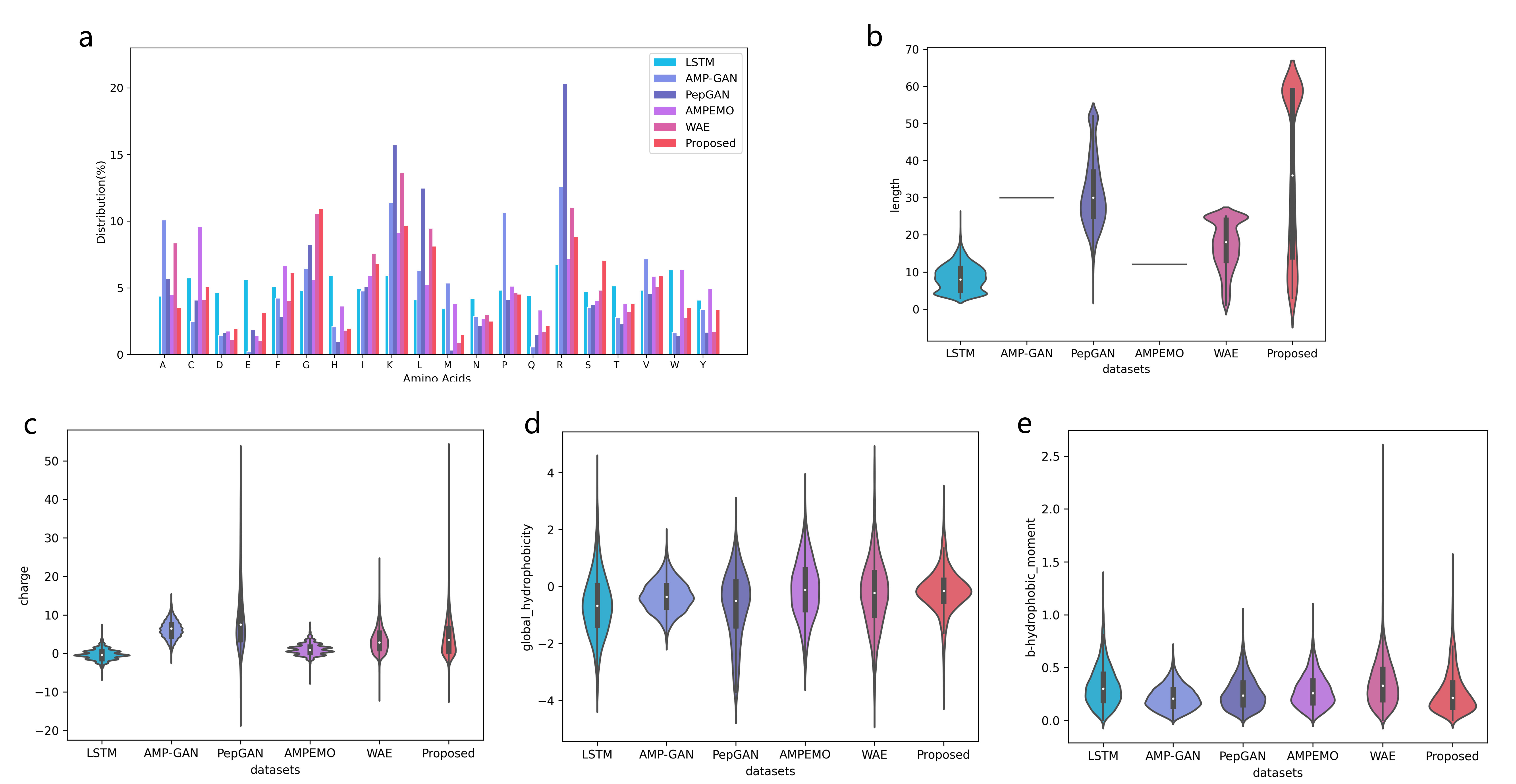}
\caption{a-e. The distribution of amino acids, sequence length, charge, hydrophobicity, and hydrophobic momentum of AMPs generated by five benchmark methods and AMPs generated by HMAMP.}
\label{S4}
\end{figure*}
\clearpage 

\begin{figure*}[htbp]
\centering 
\includegraphics[width=\textwidth]{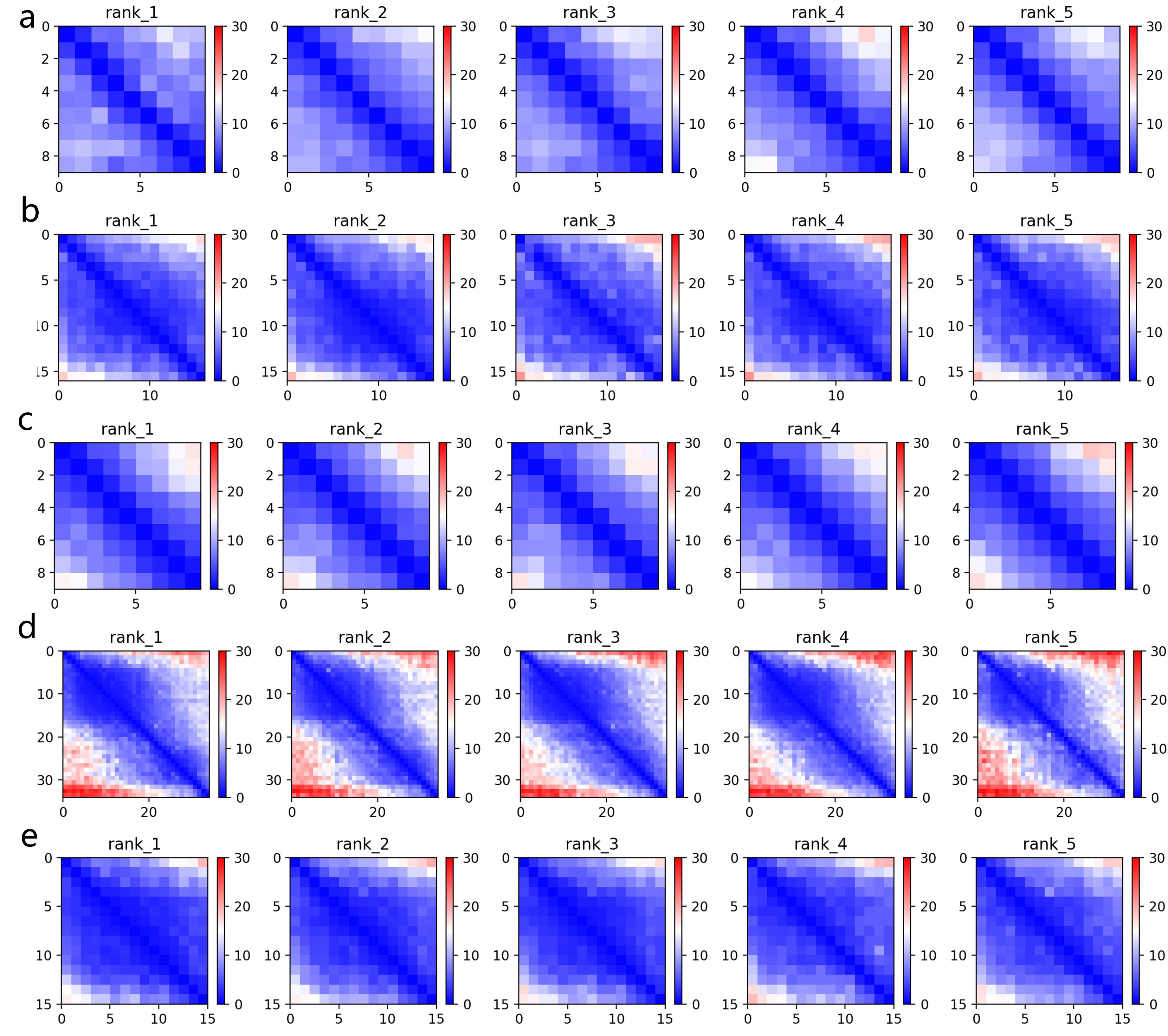}
\caption{a-e. Predicted Alignment Error maps (PAE) for ID1-ID5. }
\label{S5}
\end{figure*}
\clearpage 

\begin{figure*}[htbp]
\centering 
\includegraphics[width=\textwidth]{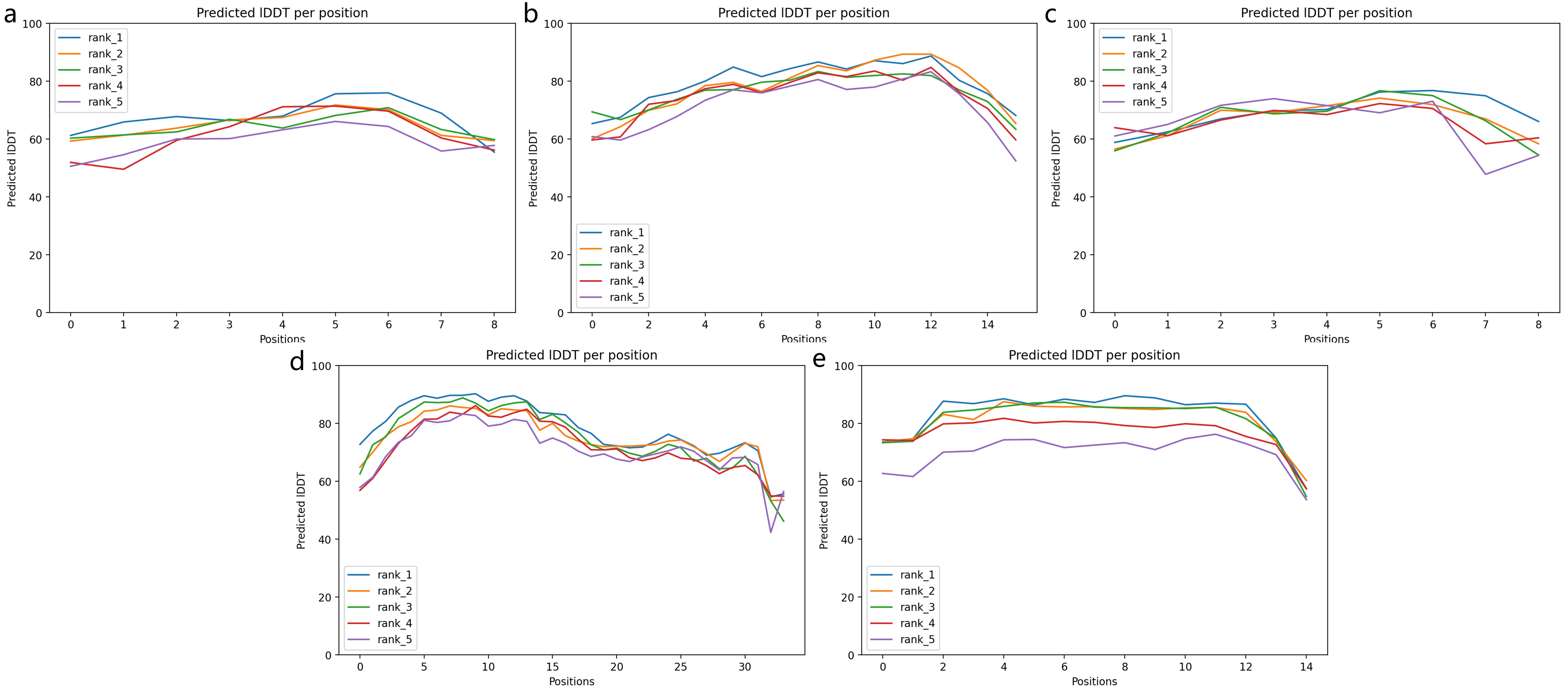}
\caption{a-e. Prediction Interface Distance Difference Test (PIDDT) for ID1-ID5.}
\label{S6}
\end{figure*}
\clearpage 

\begin{figure*}[htbp]
\centering 
\includegraphics[width=\textwidth]{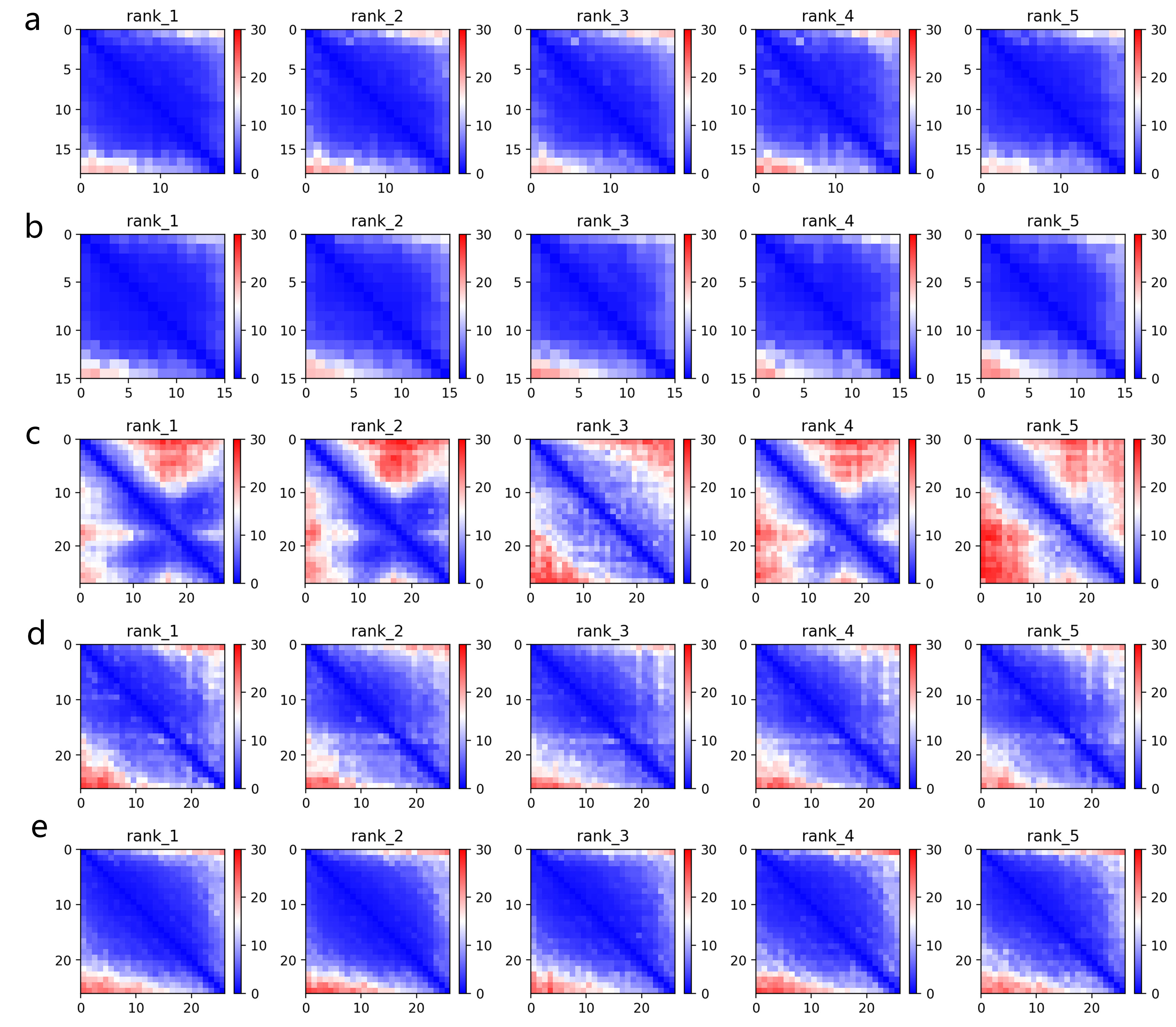}
\caption{a-e. Predicted Alignment Error maps (PAE) for ID6-ID10.}
\label{S7}
\end{figure*}
\clearpage 

\begin{figure*}[htbp]
\centering 
\includegraphics[width=\textwidth]{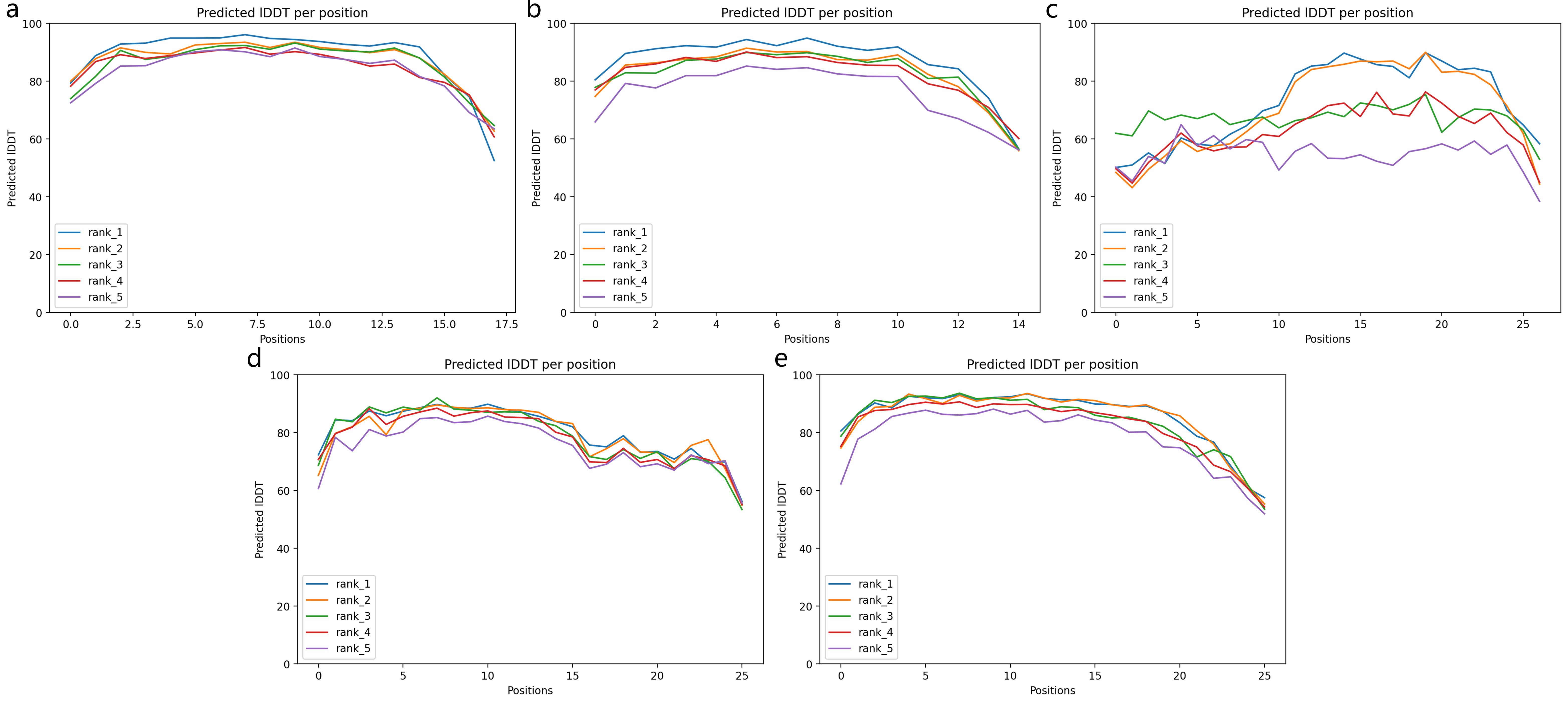}
\caption{a-e. Prediction Interface Distance Difference Test (PIDDT) plots for ID6-ID10.}
\label{S8}
\end{figure*}
\clearpage 

\begin{figure*}[htbp]
\centering 
\includegraphics[width=\textwidth]{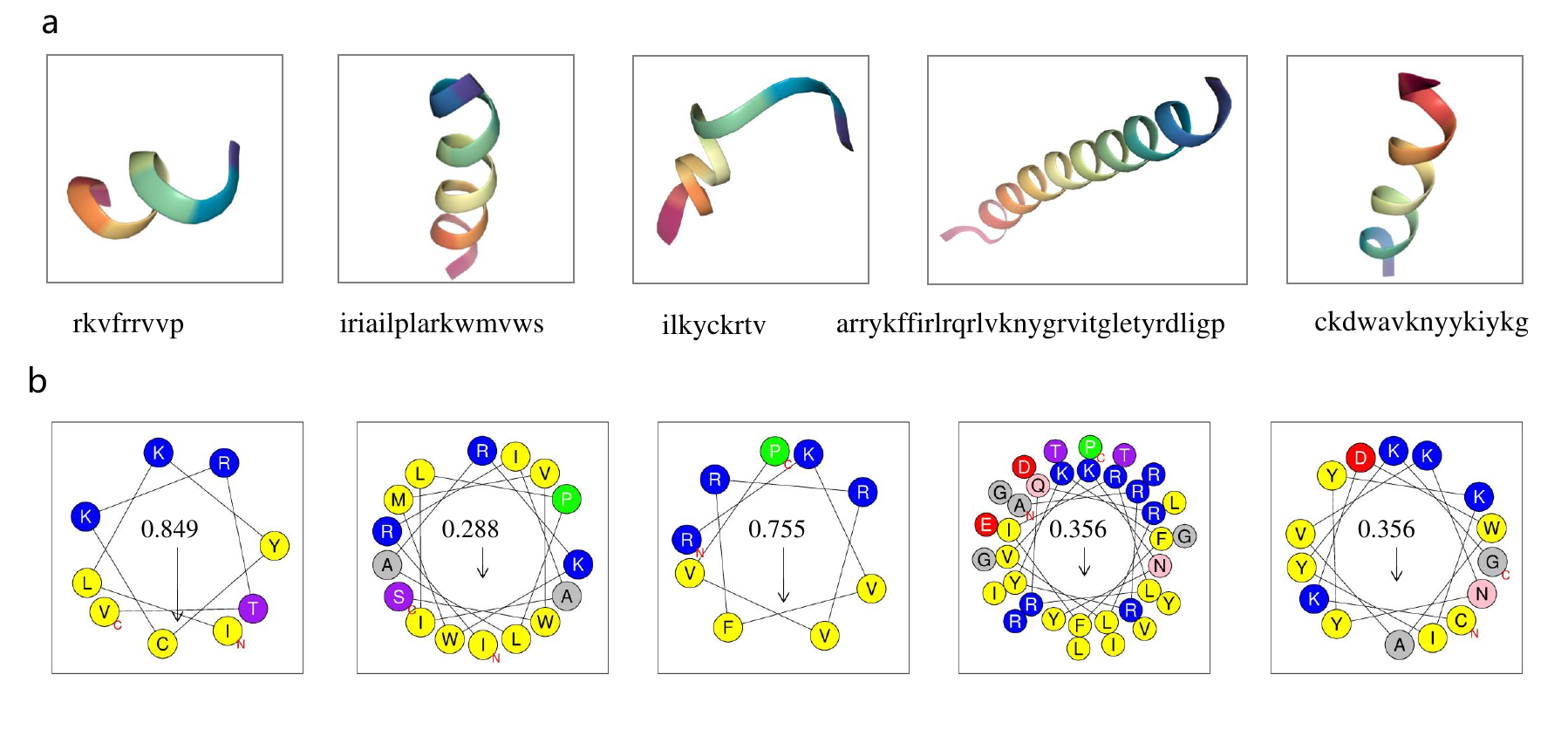}
\caption{a. Peptide structures according to MD simulations for ID1-ID5. The corresponding sequence is provided below the structures. b. Helical wheel plots for ID1-ID5. The numbers and corresponding arrows indicate the hydrophobic moment for the idealized $\alpha$-helical conformation.}
\label{second struc}
\label{S9}
\end{figure*}
\clearpage 

\begin{figure*}[htbp]
\centering 
\includegraphics[width=\textwidth]{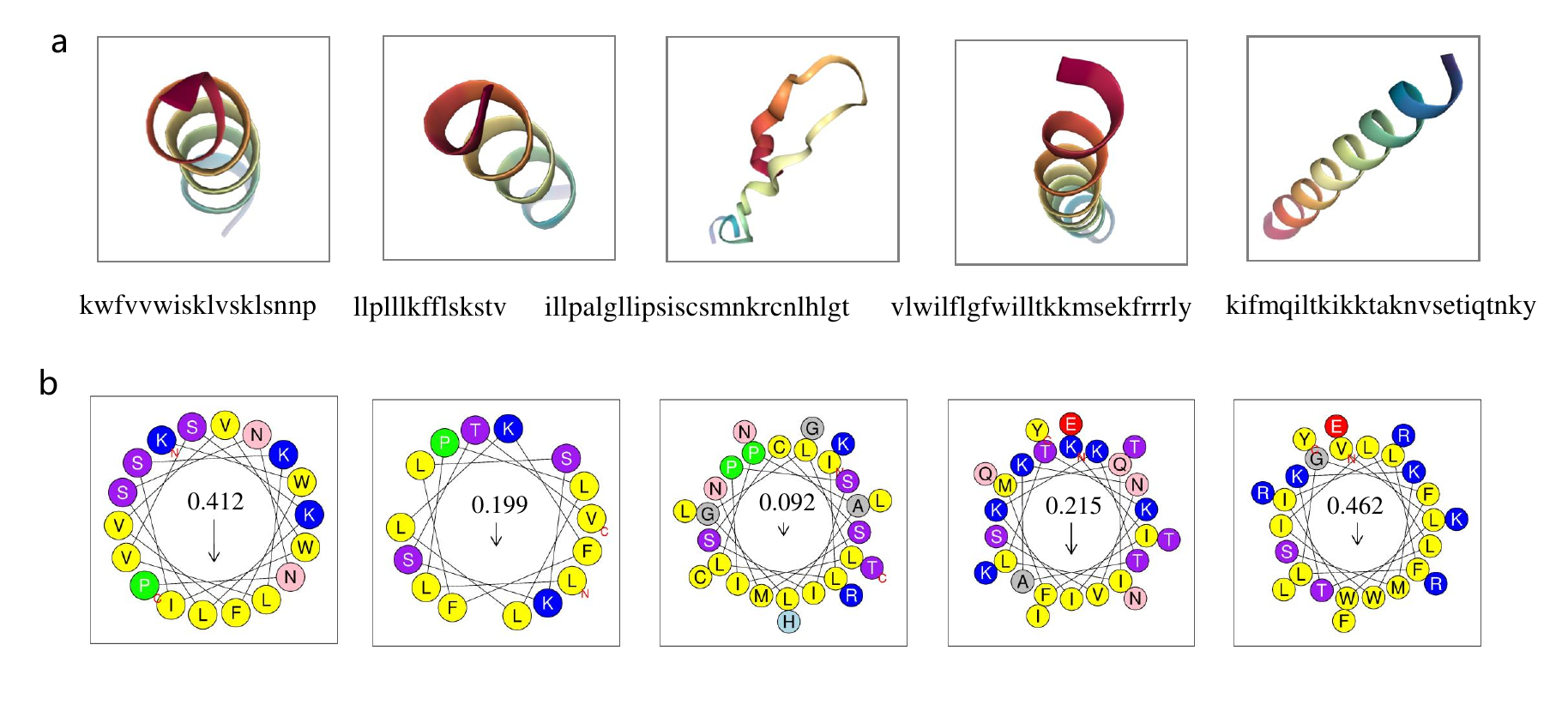}
\caption{a. Peptide structures according to MD simulations for ID6-ID10. The corresponding sequence is provided below the structures. b. Helical wheel plots for ID6-ID10. The numbers and corresponding arrows indicate the hydrophobic moment for the idealized $\alpha$-helical conformation.}
\label{S10}
\end{figure*}
\clearpage 

\bibliographystyle{IEEEtran}
\small\bibliography{HMAMP}